\documentclass[graybox]{svmult}
\usepackage{natbib}

\usepackage{mathptmx}       
\usepackage{helvet}         
\usepackage{courier}        
\usepackage{type1cm}        
\usepackage{amsmath, amssymb}
\usepackage[normalem]{ulem}
\usepackage{makeidx}         
\usepackage{graphicx}        
\usepackage{multicol}        
\usepackage[bottom]{footmisc}

\def\ni56{$^{56}$Ni}
\def\co56{$^{56}$Co}
\def\kms{km s$^{-1}$}
\def\msun{$M_\odot$}
\newcommand{\e}[1]{\times 10^{#1}}

\makeindex             

\begin{document}

                       \tableofcontents

\title*{Spectra of supernovae in the nebular phase}
\author{Anders Jerkstrand}
\institute{Anders Jerkstrand \at Queen's University Belfast, UK / Max-Planck Institute for Astrophysics, Garching, Germany, \email{anders@mpa-garching.mpg.de}}
%
%
\maketitle

\abstract{When supernovae enter the nebular phase after a few months, they reveal spectral fingerprints of their deep interiors, glowing by radioactivity produced in the explosion. We are given a unique opportunity to see what an exploded star looks like inside. The line profiles and luminosities encode information about physical conditions, explosive and hydrostatic nucleosynthesis, and ejecta morphology, which link to the progenitor properties and the explosion mechanism. Here, the fundamental properties of spectral formation of supernovae in the nebular phase are reviewed.  The formalism between ejecta morphology and line profile shapes is derived, including effects of scattering and absorption. Line luminosity expressions are derived in various physical limits, with examples of applications from the literature. The physical processes at work in the supernova ejecta, including gamma-ray deposition, non-thermal electron degradation, ionization and excitation, and radiative transfer are described and linked to the computation and application of advanced spectral models. Some of the results derived so far from nebular-phase supernova analysis are discussed.}

\index{Nebular phase}
\section{Introduction}
\label{sec:1}
As the supernova expands, reduced column densities reduce the optical depths.
Recombination removes free electrons, which further reduces the Thomson opacity. 
Decreasing temperatures lead to lower populations of excited states, reducing 
the number of optically thick lines and bound-free continua. 
After a few months, the nebula becomes mostly optically thin and its deep interior becomes visible. It then joins the class of \emph{emission line nebulae},
which includes also HII regions, planetary nebulae, and Active Galactic Nuclei. The spectrum changes from having a blackbody character
with atmospheric absorption lines imposed to an emission line spectrum rich in spectral fingerprints from the newly synthesized elements. The supernova continues to shine due to radioactive decay of
isotopes such as $^{56}$Ni produced in the explosion.
Eventually the supernova enters the ``supernova remnant'' phase. There is no generally agreed definition
of when this occurs, but the term ``remnant'' usually refers to spatially resolved supernovae of
age $\sim 10^2-10^4$ years where powering occurs by circumstellar interaction or
a central pulsar (as in the case of the Crab).
By ``nebular phase'' we usually refer to epochs from a few months to a few years.
Most supernovae become too faint to be observed after this time, unless they enter into a phase of strong circumstellar interaction, which brings them towards a remnant phase.

By studying nebular-phase supernovae we can learn about many important properties
of the exploded star. The late-time light curves provide constraints on the amount
and distribution of radioactive isotopes created in the explosion.
The spectral line strengths allows inferrence of ionic masses, emitting volumes, and physical conditions. The line profiles give information on the expansion velocity, the morphology and mixing
of the ejecta, and dust formation. Putting all this information together gives us an opportunity to
determine the properties of the progenitor stars, test stellar evolution and nucleosynthesis theory, put constraints on the explosion mechanism, and improve our understanding of the formation of black holes and neutron stars.

The modelling and interpretation of nebular-phase supernova spectra are, however, formidable challenges. 
Complexities include heterogeneous composition throughout the nebula, fast and differential velocity field, non-thermal processes, and a Non Local Thermodynamic Equilibrium (NLTE) gas state. This means that there is a long way to go from having an observed spectrum to inferring physical properties of the ejecta.


This text is written with the aims to explain the basic aspects of nebular-phase line formation, demonstrate the use of simple models and analytic methods, provide guidance to the ingredients and application of advanced models, and to review of some of the results obtained so far. We begin in Section \ref{sec:lineprofiles} by studying how line profiles are formed in the expanding nebula. In Section \ref{sec:linelum} we study the connection between physical conditions and line luminosities. Section \ref{sec:2} reviews how powering occurs in the typical
scenario of a radioactive energy source, and Section \ref{sec:modelling} reviews how physical conditions are calculated
once the powering situation is known. Section \ref{sec:examples} serves to review the availability of
advanced models, and to illustrate some output and results of these.

\newpage
\section{Line profiles}
\index{Line profiles}
\label{sec:lineprofiles}
By the time the supernova enters the nebular phase, it has reached homologous expansion (purely radial velocities with $V = r/t$).  The line broadening due to the expansion (typically a few 1000 \kms) is about three orders of magnitude larger than the line broadening due to thermal motions ($\Delta V_0 = \sqrt{kT/m}$, a few \kms~for the atoms), and the line profiles are therefore determined by the velocity structure of the nebula, but not by its temperature.

\begin{figure}[b]
\sidecaption
\includegraphics[scale=.65]{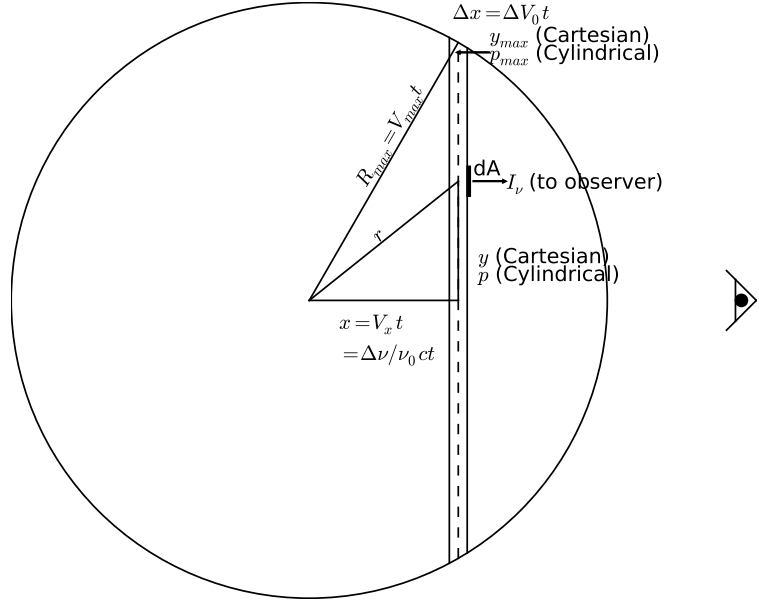} 
\caption{Geometry of line formation.}
\label{fig:sphere}
\end{figure}

Figure \ref{fig:sphere} illustrates the situation. Here the supernova is represented as a homologously expanding sphere with maximum velocity $V_{max}$. Let the $x$-axis be along the line of sight, and the $y$ and $z$ axes be perpendicular (z into the paper). Define $\Delta \nu = \nu - \nu_0$, where $\nu_0$ is the line rest frame centre frequency, and $\nu$ is the observed frequency. The observed flux at frequency $\nu$ has contribution from emission in the sheet perpendicular to the line of sight centred at projected velocity $V_x = \Delta \nu/\nu_0 c$ (distance $x=V_x t$ from the centre) having thickness $\Delta V_x = \Delta x / t$ equal to the intrinsic line width $\Delta V_0$ (which we assume is constant for now). Because the thermal line widths are small ($\Delta V_0 \ll V_{max}$), the observed line profile provides a ``scan'' through the nebula, each frequency giving a 2D integration of the emission from the sheet at the corresponding resonance depth.

From a given surface segment of the sheet, with area $dA$, the observed flux is $dF_\nu = I_\nu d\omega$ where $I_\nu$ is the specific intensity in the direction of the observer and $d\omega = 4\pi dA/(4 \pi D^2) = dA/D^2$ is the solid angle subtended by the segment $dA$ as seen by the observer at distance $D$. Integrating over segments, the total flux is
\begin{equation}
F_\nu = D^{-2} \int_{z=-\infty}^{+\infty} \int_{y=-\infty}^{+\infty} I_\nu(x(\nu),y,z) dy dz
\label{eq:generic}
\end{equation}

\subsection{Spherical symmetry}
If we assume spherical symmetry, it is enough to use a single perpendicular (cylindrical) coordinate, which we denote as $p$. Each annulus has area $dA = 2 \pi p dp$, so $d\omega = 2 \pi p dp/D^2$, giving
\begin{equation}
F_\nu = 2\pi D^{-2} \int_0^{p_{max}(x)} I_\nu(x(\nu),p) p dp
\label{eq:ss}
\end{equation}
where $p_{max} = \left(R_{max}^2-x^2\right)^{1/2}$, where $R_{max}$ is the outer radius of the nebula.
Because $p = \left(r^2 - x^2\right)^{1/2}$, $pdp = rdr$ and 

\begin{equation}
F_\nu = 2\pi D^{-2} \int_{r=x(\nu)}^{R_{max}} I_\nu(r,x) r dr
\end{equation}

The specific intensity $I_\nu$ is obtained by solving the transfer equation $dI_\nu = j_\nu ds - \alpha_\nu I_\nu ds$ through the sheet, where $j_\nu$ and $\alpha_\nu$ are emission and absorption coefficients. For an optically thin line (the optically thick case gives the same solution with a suitable choice of emissivity, see later), using that $\Delta V_0 / V_{max} \ll 1$ (so $r$ is kept fixed in the integrand)
\begin{equation}
I_\nu(r,x+\Delta x) = \int_{x-\Delta x}^{x+\Delta x} j_\nu(r,x') dx'
\end{equation}
In the comoving frame the emissivity is $j_\nu(r) = j_0(r) \phi(r,\nu-\nu_0)$, where $\phi(r,\nu-\nu_0)$ is the intrinsic line profile, normalized so that $\int_{-\Delta \nu_0/2}^{+\Delta \nu_0/2} \phi(r,\nu-\nu_0) d\nu = 1$ (in our heuristic picture the line profile is a box but the results hold generally). In the observer frame, ignoring $v/c$ effects on the intensity, $j_\nu(r,x) = j_0(r) \phi(r,\nu'-\nu_0)$, where the comoving frequency $\nu'=\nu\left(1-V_x/c\right)$. Because $d x = d V_x t = \left(d \nu/\nu_0\right) ct$, the integral is
\begin{equation}
I_\nu(r,x+\Delta x) = \frac{ct}{\nu_0} \int_{-\Delta \nu_0/2}^{+\Delta \nu_0/2}  j_0(r) \phi(r,\nu'-\nu_0) d\nu' =  \frac{ct}{\nu_0} j_0(r)
\end{equation}
Finally, the line profile is
\begin{equation}
F_\nu = 2\pi D^{-2}\frac{ct}{\nu_0} \int_{r=x(\nu)}^{R_{max}} j_0(r) r dr
\end{equation}
Or, using $r = Vt$, we can also write this as
\begin{equation}
\boxed{F_\nu = 2\pi t^2 D^{-2}\frac{ct}{\nu_0} \int_{V(\nu)}^{V_{max}} j_0(V) V dV}
\label{eq:lp}
\end{equation}




One may in principle attempt to determine $j_0(V)$ by discretizing this equation and fitting a least-squares solution to an observed line profile.
\citet{Fransson1989} suggests a variant of this, where $j_0(V)$ at a given $V$ is directly estimated from the derivative of the line profile
\begin{equation}
\frac{dF_\nu(V)}{dV} = 2 \pi t^2 D^{-2} \frac{ct}{\nu_0} j_0(V) V
\end{equation}

So far, this inverse mapping method has not been applied much in the literature.
A desirable goal is to obtain the density distribution $\rho(V)$ of the emitting ion. But connecting $j_0(V)$ to $\rho(V)$ requires knowing other functions such as temperature $T(V)$ and electron density $n_e(V)$. 



For an explicit calculation of the line profile from the gas state, one needs to insert the expression for $j_0$. It is given by
\index{emissivity}
\begin{equation}
j_0 = \frac{1}{4\pi} n_u A \beta_S h\nu_0
\end{equation}
where $n_u$ is the number density of the upper levels, $A$ is the radiative decay rate, and $\beta_S$ is the local escape probability (see Sect. \ref{sec:sobolev}), which allows also optically thick lines to be treated in this formalism.

Some limiting cases for line profiles are now derived, with illustrations in Fig. \ref{fig:lineprofiles}.

\subsubsection{Uniform sphere}
With $j_0(V) = \mbox{constant}$, Eq. \ref{eq:lp} becomes (using interchangeably $\nu-\nu_0 = \nu_0 V/c$)
\begin{equation}
F_\nu = \pi t^2 D^{-2} \frac{ct}{\nu_0} V_{max}^2 j_0 \left[1 - \left(\frac{V}{V_{max}}\right)^2\right]
\end{equation}
The line profile is parabolic in shape. The FWHM of this profile is $\sqrt{2} \times V_{max}$. We can easily understand
the parabolic shape as arising from the parabolic function describing the area of the resonance sheets.

\subsubsection{Gaussian profile}
With
\begin{equation}
j_0(V) = j_{max} \exp{\left(\frac{-V^2}{2V_0^2}\right)}
\end{equation}
we get
\begin{eqnarray}
\nonumber F_\nu = 2\pi t^2 D^{-2} \frac{ct}{\nu_0} j_{max} \int_{V(\nu)}^{\infty}\exp{\left(\frac{-V^2}{2V_0^2}\right)} V dV\\
=  2\pi t^2 D^{-2} \frac{ct}{\nu_0} j_{max} V_0^2 \exp{\left(\frac{-V^2}{2V_0^2}\right)}
\end{eqnarray}
The line profile is also Gaussian with a FWHM corresponding to the FWHM of the emissivity function (=2.35 $V_0$).

\subsubsection{Thin shell}
\index{Thin shell}
With a thin shell, $j_0(V) = j_0$, between $V_{max}-\Delta V$ and $V_{max}$, we get
\begin{eqnarray}
F_\nu = 2 \pi t^2 D^{-2}\frac{ct}{\nu_0} j_0 \frac{1}{2}\left[V_{max}^2 - \left(V_{max} - \Delta V\right)^2\right] \approx \\
2 \pi t^2 D^{-2}\frac{ct}{\nu_0} j_0 V_{max} \Delta V = \mbox{constant}
\end{eqnarray}
The line profile is a box, bounded by $\pm V_{max}$.


\subsubsection{Thick shell}
Let the inner edge be $V_{in}$.
For $V < V_{in}$, the lower integration limit is set by $V_{in}$ and the flux is therefore independent of $V$ and constant. For $V_{in} < V < V_{max}$, the solution is the same as the uniform sphere case. 
\begin{eqnarray}
F_\nu &=&  \pi t^2 D^{-2} \frac{ct}{\nu_0} j_0 V_{max}^2 \left[1 - \left(\frac{V_{in}}{V_{max}}\right)^2\right], V < V_{in}\\
F_\nu &=& \pi t^2 D^{-2} \frac{ct}{\nu_0} j_0 V_{max}^2 \left[1 - \left(\frac{V}{V_{max}}\right)^2\right], V > V_{in}
 \end{eqnarray}
The line profile is flat topped with parabolic wings.

\subsection{Asymmetric distributions}
We now study a few non-spherically symmetric configurations, starting with disks. For the disks, the discussion is based on an edge-on viewing angle, but the line profiles will keep their shapes, squeezed in width, for other viewing angles.

\subsubsection{Uniform disk}
\index{Disk}
For a uniform disk (consider Fig. \ref{fig:sphere} as now showing a face-on disk with the observer edge-on to the right), the sheet area equals the thickness of the disk $\Delta z$ times the distance between the inner and outer edges ($2\sqrt{R_{max}^2-x^2}$). Specifically, we get 

\begin{equation}
F_\nu = 2 D^{-2} \frac{ct}{\nu_0} j_0 \Delta z y_{max}(\nu) = 2 t^2 D^{-2} \frac{ct}{\nu_0} j_0 V_z V_{max} \sqrt{1 - \left(\frac{V}{V_{max}}\right)^2}
\end{equation}

This profile is less sharply peaked than the uniform sphere or Gaussian distributions.


\subsubsection{Disk with hole}
In this case
\begin{equation}
F_\nu = D^{-2} \Delta z 2 \int_{ymin(x)}^{ymax(x)} I_\nu(x(\nu),y) dy
\end{equation}
where $y_{max}(x) = \sqrt{R_{max}^2 - x^2}$ and $y_{min}(x) = \sqrt{R_{hole}^2 - x^2}$ for $x<h$, where $h=R_{hole}/R_{max}$ is the normalized hole radius, and $y_{min}=0$ otherwise. Then
\begin{eqnarray}
F_\nu = 2t^2D^{-2} \frac{ct}{\nu_0} j_0 V_z V_{max} \times \\
\left[\sqrt{1-\left(\frac{V}{V_{max}}\right)^2} - \sqrt{h^2-\left(\frac{V}{V_{max}}\right)^2}\right]~~~,\left(\frac{V}{V_{max}}\right) < h\\
\sqrt{1-\left(\frac{V}{V_{max}}\right)^2}~~~,\left(\frac{V}{V_{max}}\right) > h
\end{eqnarray}

The high velocities have the same solution as the uniform disk. Once the hole is crossed ($V < V_{max}h$), the projected area is $\sqrt{1-\left(\frac{V}{V_{max}}\right)^2} - \sqrt{h^2-\left(\frac{V}{V_{max}}\right)^2}$. The second term grows faster with decreasing $V/V_{max}$, and the line profile develops two horns at plus and minus the velocity of the inner edge of the disk. 

\begin{figure}[htb]
\sidecaption
\centering
\includegraphics[width=1\linewidth]{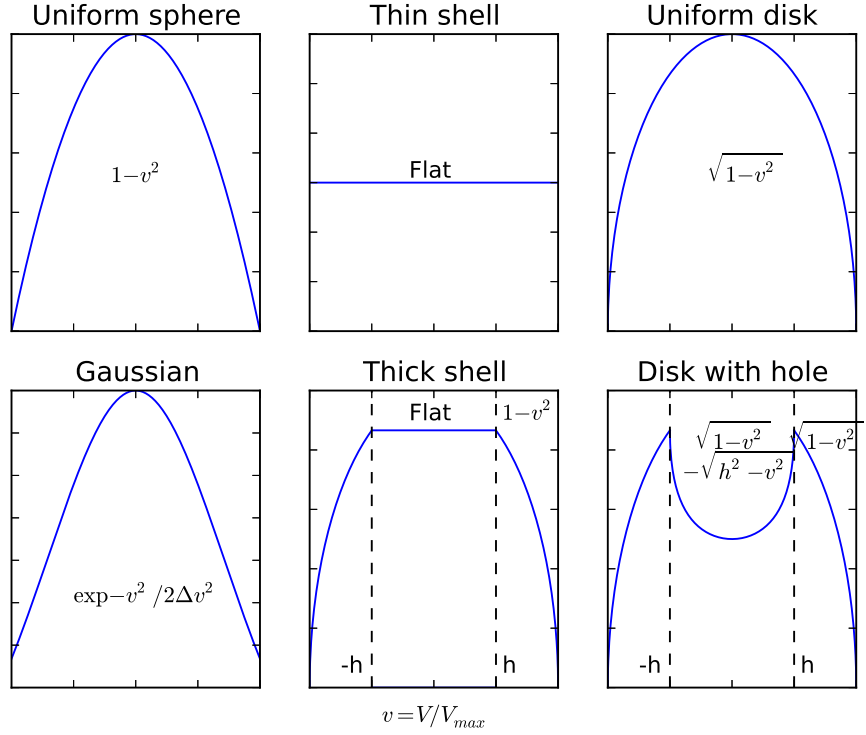}
\caption{Line profiles ($F_\nu$) for six types of ejecta distributions.}
\label{fig:lineprofiles}
\end{figure}

Figure \ref{fig:lineprofiles} summarizes the various line profiles discussed.

\subsubsection{Many clumps}
If the emission comes from a large number of clumps with a more or less random distribution within a spherically symmetric or axisymmetric region, the line profiles will have a global shape determined by the equations above (using the probability distribution as $j_0(V)$), but with small-scale structure determined by the random positions of the clumps. 

\begin{figure}[htb]
\sidecaption
\centering
\includegraphics[width=0.49\linewidth]{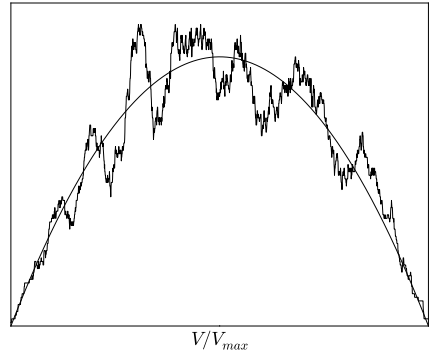}
\caption{Line profile resulting from random draws of $N=10^3$ clumps
and $\epsilon=0.05$, distributed in a uniform spherical region.}
\label{fig:random}
\end{figure}

Assume that we have $N$ identical clumps with (comoving) expansion velocities $V_{c}$ (their 'width' is $V_c$), distributed randomly within a sphere of velocity $V_{max}$. The resulting line profile will have squiggles with statistical properties depending on $N$ and $\epsilon=V_{c}/V_{max}$. Figure \ref{fig:random} shows the result of simulating a line profile with $N=10^3$ and $\epsilon=0.05$. A quantitative method to use the statistical properties of these fluctuations to infer the statistical properties of the 
clump distribution was developed by \citet{Chugai1994}. Its deployment needs high-resolution spectroscopy, and is therefore suitable for the most nearby supernovae.

\subsubsection{Comments}
A final few comments are in place. We have derived line profile shapes neglecting effects of time delays and relativistic effects, apart from Doppler shifts.
Adressing the first point, photons arriving from the receding side of the SN were emitted a time $t_{delay} \sim 2 V_{max}t /c$ before photons from the approaching side. If there is an evolution of emissivity $j_0$ on the evolutionary time-scale $t$, there is therefore a damping of the red side of the line compared to the blue of order $2V_{max}/c$, which is a few percent for typical $V_{max}$.
But if evolution occurs on a faster radioactive time-scale, the effect is increased by a factor $t_{delay}/\tau_{decay}$.
\begin{equation}
e^{\frac{-t_{delay}}{\tau_{decay}}}= e^{\frac{-2 V_{max}t/c}{\tau_{decay}}}
\end{equation}
For example, if $t=500d$ and $\tau_{decay}=111d$ (as for $^{56}$Co), then $t/\tau_{decay} = 4.5$, and if $V_{max}=5000$ km s$^{-1}$, the factor becomes 15\%.
It may be incresed further if gamma-ray leakage further shortens the time-scale over which emissivity changes.

The source movement also increases the blue intensity as the observed time interval is shorter than in the emitting frame, and vice versa decreases the red side. The Lorentz transformations of specific intensity gives a $\left(\nu/\nu'\right)^3$ factor, or a change of order $\sim 3 V_{max}/c$ in our non-relativistic limit.
With $V_{max}/c$ of a few percent, the total additional effect is of order 5-10\%.

\subsection{Radiative transfer effects}

So far we have assumed that apart from self-absorption (which can be treated as a modification to the emissivity $j_0$ in the Sobolev approximation, see Sect. \ref{sec:sobolev}), the photons escape freely the nebula. For any given epoch, this is a reasonable approximation beyond some wavelength. At short wavelengths, various opacities remain, however, for years or decades and can alter both line profiles and luminosities. 


\subsubsection{Continuous scattering opacity}
\index{Scattering}
\index{Electron scattering}

Continuous scattering may occur by free electrons or dust. This leads to one or several ``bounces'' for the photons in the homologous (Hubble) flow. 
Because the comoving frame wavelength is always lower than in the original emitting frame, 
there is on average a net energy loss, and the line profile
becomes distorted with an enhanced red tail. 

\begin{figure}[htb]
\centering
\includegraphics[width=0.8\linewidth]{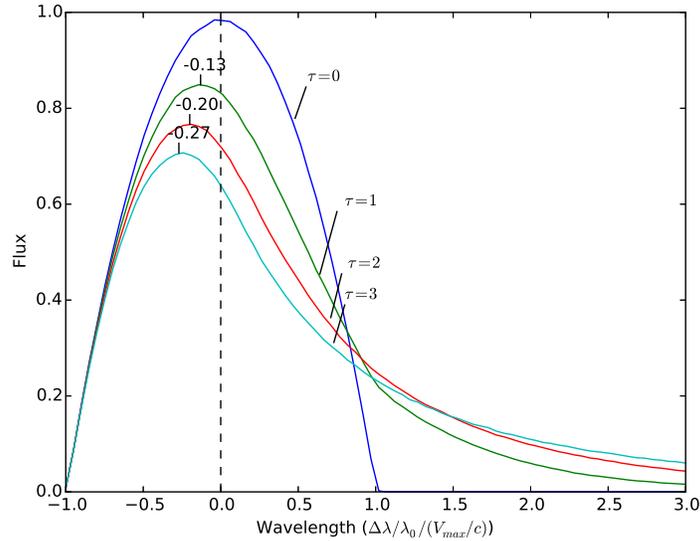} 
\caption{Line profiles resulting from a scattering opacity (e.g. electron or dust scattering) in a uniform sphere. The wavelength shifts of the peaks are written out.}
\label{fig:electronscattering}
\end{figure}

In the nebular phase the electron scattering optical depth is $\tau_e \lesssim 1$, and the majority of photons will experience zero or one scattering events. The line profile distortions are thus relatively mild. Figure \ref{fig:electronscattering} shows the resulting line profiles for $\tau_e=1,2,3$ in a uniform sphere setup, from a Monte Carlo simulation.
The scattering opacity produces blueshifts of the peak, although quite weak. For example, at $\tau_e=1$,  a shift of $\Delta \lambda/\Delta \lambda_{max}=-0.13$ is obtained (or 390 km s$^{-1}$ for a 3000 km s$^{-1}$ broad line).


If one considers also the frequency redistribution in the comoving frame due to the thermal motions of the scattering particles, a further symmetric broadening on the scale $\sqrt{kT/m}$ occurs as well. Only electrons provide any significant thermal effect due to their low mass. Still, at 5000 K the thermal electron velocity is just 300 km s$^{-1}$, much lower than the supernova expansion velocity, so this effect can be ignored unless very detailed results are needed.

\begin{figure}[htb]
\centering
\includegraphics[width=0.8\linewidth]{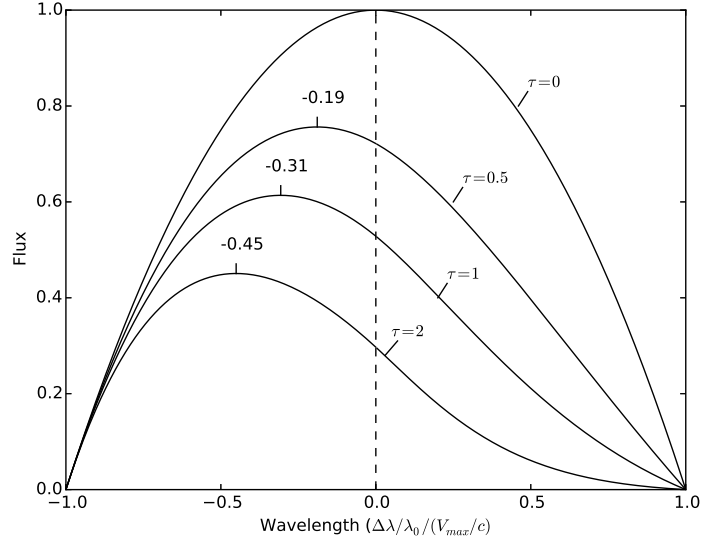} 
\caption{Line profiles resulting from a destructive opacity, uniform sphere case.}
\label{fig:dustprofiles}
\end{figure}

\subsubsection{Continuous absorptive opacity}
\index{Absorption}
\index{Dust absorption}
Continuous absorption (photon destruction) may occur by photoionization or dust. At long wavelengths also free-free absorption may occur.
Assume that an absorptive opacity is present with an absorption coefficient $\alpha_{\nu}$ (cm$^{-1}$). Consider for example purely absorbing dust, that reemits at mid-infrared (MIR) wavelengths beyond our consideration. The emergent flux is, by extension of Eq. \ref{eq:ss}:
\begin{equation}
F_\nu = 2\pi D^{-2} \int_0^{pmax(x)} I_\nu(x(\nu),p)e^{-\tau(x(\nu),p)} p dp 
\end{equation}
The optical depth is
\begin{equation}
\tau(x,p) = \alpha_\nu \times \left(\sqrt{R^2 - p^2} - x\right)
\end{equation}
where we now denote $R=R_{max}$, and so
\begin{equation}
F_\nu = 2\pi D^{-2} \frac{ct}{\nu_0} \int_0^{pmax(x)} j_0(r(x,p)) e^{-\alpha_\nu\left(-x + \sqrt{R^2 - p^2}\right)} p dp
\end{equation}
For constant emissivity,
\begin{equation}
F_\nu = 2\pi D^{-2} \frac{ct}{\nu_0} j_0 \int_0^{pmax(x)} e^{-\alpha_\nu\left(-x + \sqrt{R^2 - p^2}\right)} p dp
\end{equation}
Substitute $x' = \sqrt{R^2 - p^2}$, so $p = \sqrt{R^2 - x'^2}$, $x'_{min} = R$ and $x'_{max} = \sqrt{R^2 - p_{max}^2} = x$. 
This last equality holds for positive $x$ only. Then $x' dx' = -2pdp$ and
\begin{eqnarray}
F_\nu = 2\pi D^{-2} \frac{ct}{\nu_0} j_0 \int_{x'=R}^{x} e^{-\alpha_\nu\left(-x + x'\right)}\times -\frac{1}{2}x'dx' \\
= \pi D^{-2} \frac{ct}{\nu_0} j_0  \frac{1}{\alpha_\nu^2} \left[e^{-\tau_0\left(1-\hat{x}\right)}\left(-\tau_0 - 1\right) + \tau_0 \hat{x} + 1\right] 
\end{eqnarray}
where $\tau_0 = \alpha_\nu R$ and we have denoted $\hat{x} = \Delta \nu / \Delta \nu_{max}$.

For negative $x$, use instead $x' = -\sqrt{R^2 - p^2}$. Then $p= \sqrt{R^2-x'^2}$, $x_{min}' = -R$ and $x_{max}' = -|x| = x$, $x'dx' = -2pdp$, and
\begin{eqnarray}
F_\nu = 2\pi D^{-2} \frac{ct}{\nu_0} j_0 \int_{x'=-R}^{x} e^{-\alpha_\nu\left(-x - x'\right)}\times -\frac{1}{2}x'dx' \\
= \pi D^{-2} \frac{ct}{\nu_0} j_0  \frac{1}{\alpha_\nu^2} \left[e^{-\tau_0\left(1-\hat{x}\right)}\left(-\tau_0 - 1\right) - e^{2\tau_0 \hat{x}}\left(\tau_0 \hat{x} - 1\right)\right] 
\label{eq:dd}
\end{eqnarray}


These expressions can be used to fit line profiles affected by a destructive opacity to estimate $\tau_0$. 
The line profiles for $\tau_0=0.5, 1$ and 2 are plotted in Fig. \ref{fig:dustprofiles}. There are two main differences to scattering opacities. First, there is no production of a red tail with $\Delta \lambda/\Delta \lambda_{max} > 1$. Second, for a given $\tau_0$, the peaks are more strongly blueshifted. The location of the peak is found from equating the derivative of Eq. \ref{eq:dd} to zero, giving
\begin{equation}
\hat{x}_{peak} = 1 - \frac{\ln{\left(1+\tau_0\right)}}{\tau_0}
\end{equation}

\subsubsection{Line absorption}

If line opacity can be described as a large set of finely spaced lines, its effect can be treated as a continous opacity as described in the sections above. For scattering/absorption by one or a few lines, no generic treatment is possible, and a diverse set of line profiles can be produced. A common scenario is that only the longer wavelength line in a doublet or triplet emerges, as the bluer components are absorbed.

\begin{figure}[htb]
\centering
\includegraphics[width=0.8\linewidth]{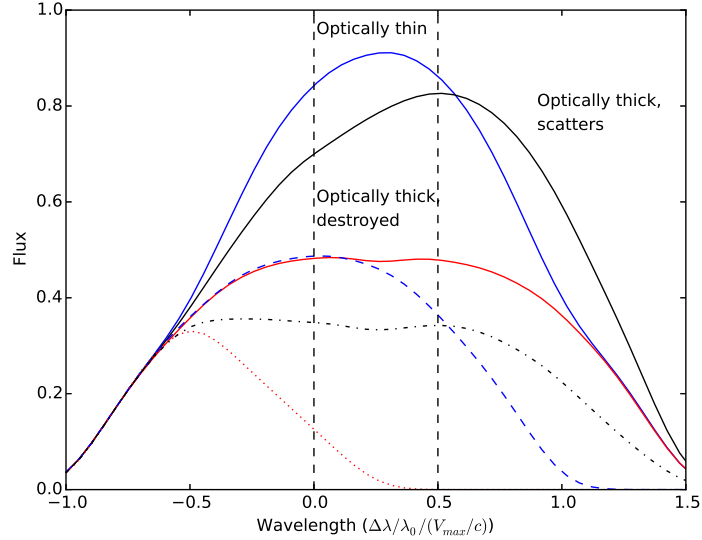} 
\caption{Line profiles resulting from a doublet separated by $\Delta \lambda/\lambda_0=0.5.$ The center wavelengths are marked by dashed vertical lines. Solid lines correspond to equal (before transfer) emissivites in both lines, and dashed (optically thin), dot-dashed (scattering) and dotted (destructive) correspond to emission in only the blue line.}
\label{fig:bbprofiles}
\end{figure}

Figure \ref{fig:bbprofiles} shows six examples of line profiles from a doublet, separated by $\hat{x}=0.5$. Here, equal emissivity in both lines gives a symmetric line in the optically thin case, a distorted line peaking close to the redder line wavelength for optically thick scattering, and again a symmetric but damped and flattened line for optically thick destructive opacity. The dashed, dot-dashed, and dotted curves show the cases when only the first (blue) line emits.

\newpage

\section{Line luminosities} 
\label{sec:linelum}
\index{Line luminosities}

In this section expressions for line luminosities in different physical limits are examined. 
Physical conditions, such as temperature and electron density, are here parameters. The processes determining these are discussed in the following sections.
As a first step we consider how to treat line transfer in the supernova.


\subsection{Line transfer in the Sobolev approximation}
\label{sec:sobolev}
\index{Sobolev approximation}
Supernovae are still dense enough in the nebular phase that line optical depths can be high. 
It is in general a difficult problem to solve the radiative transfer through optically thick lines. 
\citet{Sobolev1957} showed that a great simplification can be achieved in the \emph{high velocity gradient limit}, meaning situations where the velocity gradient of the expanding nebula is large enough that line profiles are traversed on a length scale smaller than the length scale over which physical conditions change. In the case of homologous expansion the optical depth to traverse a line in this limit is given by the Sobolev optical depth

\begin{equation}
\tau_S = \frac{1}{8\pi}\frac{g_u}{g_l}A\lambda^3 n_l \left(1-\frac{g_l}{g_u}\frac{n_u}{n_l}\right) t
\label{eq:sobolev}
\end{equation}
where $g_u$ and $g_l$ are the statistical weights of the upper and lower levels, $n_u$ and $n_l$ are the number densities, $\lambda$ is the wavelength, and $t$ is time. For photons emitted in the line, the average escape probability can be shown to be
\begin{equation}
\beta_S = \frac{1-e^{-\tau_S}}{\tau_S}
\end{equation}

These simplifications mean that we do not have to compute the detailed transfer through each line, but can treat them as infinitely narrow with optical depth and escape probability given by the expressions above. 

The total luminosity in the transition is given by the volume integral
\begin{equation}
L = \int n_u A h\nu \beta_S(n_l,n_u)  \beta d\mathcal{V}
\label{eq:basic}
\end{equation}
where $\beta$ is the angle-averaged \emph{non-local} escape probability (with respect to absorption by photoionization, other lines, dust, etc). We will for now proceed by putting $\beta = 1$, this factor can easily be added on to the final formulae if relevant.

\subsection{Local Thermodynamic Equilibrium}
\index{LTE, nebular phase}
In Local Thermodynamic Equilibrium (LTE), $n_u$ and $n_l$ are related by the temperature only.
LTE requires that both populations and depopulations of the level occur mainly by thermal collisions, or by radiative interaction with a (possibly diluted) blackbody radiation field. As a starting point, one can consider the competition between thermal collisional deexcitation, with a rate $n_e Q(T)$, where $Q(T)$ is the collision rate (cm$^3$ s$^{-1}$), and spontaneous radiative deexcitation, with a rate $A \beta_S$. The \emph{critical density} is defined as the electron density above which collisional deexcitations dominate:
\begin{equation}
n_e^{crit}(T,n_l,n_u) = \frac{A\beta_S(n_l,n_u)}{Q(T)}
\label{eq:ncrit}
\end{equation}
Note the difference from static media in that there may be a dependency on density as well as temperature.
The temperature dependence of $Q(T)$ is usually quite weak, so $n_e^{crit}$ depends only weakly on $T$. $Q(T)$ is given by (see e.g. \citep{Osterbrock2006})
\begin{equation}
Q = 8.6\e{-6} T^{-1/2} g_u^{-1} \Upsilon(T)~~~~\mbox{cm}^3 \mbox{s}^{-1}
\end{equation}
where $\Upsilon(T)$ is the \emph{effective collision strength}, which depends on the cross section function for the particular transition, but is typically of order unity and with a normally weak temperature dependency.

The electron density in a uniform sphere is
\begin{equation}
n_e =  2\e{9} \mu^{-1}\left(\frac{M}{1\ M_\odot}\right)\left(\frac{V}{3000~\mbox{km s}^{-1}}\right)^{-3} \left(\frac{x_e}{0.1}\right) \left(\frac{t}{200~\mbox{d}}\right)^{-3}\left(\frac{f}{0.1}\right)^{-1}~~~\mbox{cm}^{-3}
\label{eq:ne}
\end{equation}
where $\mu$ is the mean atomic weight, $x_e = n_e/n_{nuclei}$ is the electron fraction, and $f$ is the filling factor.
To have critical densities below typical nebular densities of $n_e \sim 10^9$ cm$^{-3}$ (i.e. LTE), Eq. \ref{eq:ncrit} with a typical value $Q\sim 10^{-7}$ cm$^3$s$^{-1}$ shows that transitions need to be forbidden/semi-forbidden ($A \lesssim 10^2$ s$^{-1}$) or effectively forbidden/semi-forbidden ($A\beta_S \lesssim 10^2$ s$^{-1}$). 

The second requirement for LTE, population by thermal collisions, 
requires the upper level (energy $E_u$) to be reachable from the thermal pool, $E_u \lesssim \mbox{a few} \times kT$. Because $T \lesssim 5000 $ K = 0.4 eV in the nebular phase, that means $E_u \lesssim $ a few eV.

The lines fulfilling both of these criteria are low-lying forbidden transitions in atoms and ions of e.g. C, N, O, Si, S, Ca, Fe, Co, and Ni.
Table \ref{table:LTElines} lists some important transitions, all clearly detected in nebular SNe (there are more transitions
fulfilling the criteria but which have not clearly identified, because they are either too weak or blended with other lines).

\begin{table}
\centering
\begin{tabular}{|c|c|c|c|c|c|c|c|c|c|}
\hline
Ion & Line ($\AA$)   & $E_u(eV)$       & $g_u$ & $A (s^{-1})$         & $\Upsilon$ & $n_e^{crit}$ (cm$^{-3}$) & $t_{LTE}$ (d) & $\epsilon_n$ & $t_{thin}(d)$\\
\hline
C I & 9850 (4-3)      & 1.26 &     5    & $2.2\e{-4}$        & 0.34        & $2.6\e{4}$ & 3700 & 0.04  & 25$\left(M/0.1\  M_\odot\right)^{1/2}$
\\
     & 8727 (5-4)     & 2.68 &       1    & 0.60              & 0.20          &  $2.5\e{7}$ & 380 &  & To exc.\\
N II & 6583 (4-3)   & 1.89      &  5   & $2.9\e{-3}$    & 1.4           & $8.5\e{4}$  & 2400 & 0.14 & 45$\left(M/0.1\ M_\odot\right)^{1/2}$\\
O I & 6300 (4-1)   & 1.97      & 5    & $5.6\e{-3}$    & 0.06        &   $3.8\e{6}$ & 640 & 0.21 & 170$\left(M/1\ M_\odot\right)^{1/2}$\ \\
      & 5577 (5-4) & 4.18         & 1    & 1.3              & 0.07         & $1.5\e{8}$ & 190 & & To exc. \\
Si I & 1.64~$\mu$m (4-3) & 0.78 & 5   & $2.7\e{-3}$    & 0.1          & $1.0\e{6}$ & 800 & 1.0 & 120$\left(M/0.1\  M_\odot\right)^{1/2}$\\
      &  1.10~$\mu$m (5-4)  & 1.91 & 1    & 0.80             & 0.02         & $3.3\e{8}$ & 120 & & To exc.\\
S I  & 1.08~$\mu$m (4-1) & 1.15  & 5   & 0.028           & 0.1          & $1.1\e{7}$ & 350 & 2.6 & 200$\left(M/0.1\  M_\odot\right)^{1/2}$\\
       & 7725 (5-4)    & 2.75     & 1    & 1.8              & 0.02          & $7.4\e{8}$ & 90 & & To exc.\\
Ca II & 7291 (3-1)  & 1.70   & 6    & 1.3              & 7.3              & $8.8\e{6}$ & 350 &162 & 480$\left(M/0.01\ M_\odot\right)^{1/2}$\\
Fe II & 7155 (17-6) &   1.96    & 10   & 0.15           & 1.0             & $1.2\e{7}$ & 290 & & To exc.\\
        & 1.26~$\mu$m (10-1) & 0.99 & 8   & $5.0\e{-3}$ & 13             & $2.6\e{4}$ & 2200 & 0.21 & 55$\left(M/0.1\ M_\odot\right)^{1/2}$\\
        & 1.64~$\mu$m (10-6) & 0.99 & 8    & $5.1\e{-3}$ & 2.2         & $1.5\e{5}$ & 1200 & & To exc.\\ 
Co II & 1.02~$\mu$m (9-1) & 1.22 & 9 & 0.054          &  0.32            & $1.2\e{7}$ & 280 & 1.9 & 160$\left(M/0.1\  M_\odot\right)^{1/2}$\\
         & 9338 (10-1) &   1.32     & 7   & 0.023         & 0.25            & $5.3\e{6}$ & 380 & 0.5 & 80$\left(M/0.1\ M_\odot\right)^{1/2}$\\
Ni II &  7378 (7-1)   & 1.68    & 8     & 0.23          & 1.2            & $1.2\e{7}$ & 280 & 5.5 & 90$\left(M/0.01\ M_\odot\right)^{1/2}$\\
\hline
\end{tabular}
\caption{Commonly observed SN lines that are expected to be in LTE for some part of the nebular phase. For doublets and triplets only one of the components is listed.
The line wavelengths (col. 2) are followed by level IDs in parenthesis by energy ordering. The effective collision strengths $\Upsilon$ (column 6), and resulting critical densities (column 7), are for $T=5000$ K and $\beta_S=1$ (if the line is optically thick $t_{LTE}$ can be longer). Column eight lists the length of time the line is in LTE, using Eq. \ref{eq:ne} ($t_{LTE}$ scales with $\left(M/1~M_\odot\right)^{1/3}\left(V/3000~\mbox{km s}^{-1}\right)^{-1} \left(x_e/0.1\right)^{1/3}\left(f/0.1\right)^{-1/3}$). The $t_{thin}$ values (tenth column) scale with $\left(V/3000~\mbox{km s}^{-1}\right)^{-3/2}\left(f/0.1\right)^{-1/2}$. The mass scalings are based on typical masses in core-collapse supernova models. Estimates for decays to excited states are not attempted, but labelled 'To exc.'. Atomic data sources are given in \citet{Jerkstrand2011}.}
\label{table:LTElines}
\end{table}


Some lines that are not in the table warrant comment. 
To populate a level mainly by thermal collisions requires not only low enough excitation energy, but also a reasonably high abundance compared to the next ionization stage; recombinations may otherwise become the dominant population mechanism.
A recombination-dominated situation often arises for neutral elements with low ionization potential, such as Na I and Mg I.
A second issue is for resonance lines. These may fulfill $A\beta_S \lesssim 10^2$ s$^{-1}$, but their high optical depth means they can become dominated by scattering. A line like Na I 5890 \AA\ for example, has both of these properties and would be poorly modelled as an LTE line. 


\subsection{Optical depth}
\index{Optical depth, nebular phase}
\label{sec:opticaldepth}
For lines connected to the ground multiplet, the optical depths can be estimated by taking $M_l \approx (g_l/g_g)M_{ion}$, where $g_g$ is the total ground multiplet statistical weight. This assumes that most atoms are in the ground multiplet, and that this is in LTE, both normally good approximations. 
Then, Eq. \ref{eq:sobolev} gives, for a uniform sphere and ignoring stimulated emission
\begin{equation}
\tau_S = 4.3\e{29}\left(\frac{g_u}{g_g}\right)\left(\frac{M_{ion}}{1\ M_\odot}\right) \mu^{-1}\left(\frac{V}{3000~\mbox{km s}^{-1}}\right)^{-3} f^{-1} t^{-2} A \lambda^3 
\end{equation}
Define $\epsilon= 4.3\e{29} \mu^{-1} g_u/g_g A \lambda^3$ (atomic constants only) and $\epsilon_n = \epsilon/10^{14}$. Then, the time at which $\tau_S=1$ is
 \begin{equation}
t_{\rm thin} =  370d~\epsilon_n^{1/2} \left(\frac{M_{ion}}{1\ M_\odot}\right)^{1/2}\left(\frac{V}{3000~\mbox{km s}^{-1}}\right)^{-3/2} \left(\frac{f}{0.1}\right)^{-1/2}
\label{eq:thin}
\end{equation}

This equation has stronger dependencies on $M_{ion}$ and $V$ than the inversion of the LTE equation (\ref{eq:ne}) has, thus, estimates for the duration of the optically thick phase are more uncertain than estimates for  the duration of the LTE phase.
Column 10 in Table \ref{table:LTElines} lists the epochs ($t_{thin}$) at which the lines become optically thin for typical masses, $V=3000$ \kms, and $f=0.1$. 

 
For lines decaying to excited states ([C I ] 8727, [O I] 5577, [Si I] 1.10 $\mu$m, [S I] 7725, [Fe II] 7155, [Fe II] 1.26 $\mu$m), transition to optical thinness will usually occur early as populations of excited states are much lower than in the ground multiplet. However, if the A-value is large, this should be more carefully checked. The transition time has in this case a strong temperature-dependency through the sensitivity of populations of excited states.

In some cases two lines from the same ion can be used to determine the optical depth and thereby the density. As an example, the [O I] 6300 and 6364 \AA~lines arise from the same upper level (2p4$^1$D), going to the first and second levels in the ground multiplet, respectively, with $A_{6300} = 5.6\e{-3}$ s$^{-1}$ and $A_{6364} = 1.8\e{-3}$ s$^{-1}$. The ratio of their emissivities $R=j_0(6300)/j_0(6364)$
is
\begin{equation}
R = \frac{A_{6300} h\nu_{6300} \beta_{S,6300}}{A_{6364} h\nu_{6364}\beta_{S,6364}}
\end{equation}
Ignoring stimulated emission, $\tau_{S,6300}/\tau_{S,6364} = \left(6300^3/6364^3\right) A_{6300}/A_{6364} n_1/n_2 g_2/g_1 = 0.97 A_{6300}/A_{6364} = 3.0$, where we have assumed LTE within the ground multiplet and $T \gg \Delta E_{ground}$ (230 K) for the ground multiplet populations, so $n_1/n_2 = g_1/g_2$. Then 
\begin{equation}
R = \frac{1-\exp{\left(-\tau_{6300}\right)}}{1-\exp{\left(-\frac{1}{3}\tau_{6300}\right)}}
\end{equation}
In the optically thick limit $R\rightarrow 1$, and in the optically thin limit $R\rightarrow 3$. 


By studying how the ratio transitions from the thick to thin regime over time, the density of O I can be estimated. Initial application for SN 1987A resulted in $\rho_{\rm OI}(t) = 1.7\e{-12}\left(t/100\mbox{d}\right)^{-3}$ g cm$^{-3}$ \citep{Spyromilio1991, Li1992}. The mass can be estimated from $M_{\rm OI} = \rho_{\rm O I} \frac{4\pi}{3}V_{max}^3 t^3 f_O$, but the filling factor $f_O$ needs to be determined by some other method.
 

\subsubsection{LTE, optically thin case}

If the whole ion is modelled in LTE, $N_u = N g_u Z(T)^{-1} e^{-E_u/kT}$, where $N$ is the total number of ions, and $Z(T)$ is the partition function. Optical thinness means $\beta_S = 1$. Then, with $N=M_{ion}/\left(\mu m_p\right)$, Eq. \ref{eq:basic} becomes
\begin{equation}
L = M_{ion} \left(\mu m_p\right)^{-1} A h\nu  \frac{g_u}{Z(T)} e^{-E_u/kT}
\label{eq:LTEthin}
\end{equation}
One may also consider a variant, where it is not assumed that the whole ion is in LTE, but only that the upper and lower states are in LTE with each other (thermal collisions dominate transitions in both directions). Then, if the lower state is in the ground multiplet, and we approximate $M_g \approx M_{ion}$, Eq. \ref{eq:LTEthin} is recovered with $Z(T)$ replaced  by $g_g$.

The mass of the emitting ion $M_{ion}$ can therefore be estimated if the temperature can be determined. The most robust mass inferrances can be made from lines with $E_u \ll kT$ (so $e^{-E_u/kT} \rightarrow 1$), 
which means $\lambda \gtrsim 3\ \mu m \left(T/5000\ K\right)^{-1}$. For temperatures of a few thousand K these are MIR lines. However, mass $\emph{ratios}$ may be robostly determined also in the $E_u \gtrsim kT$ regime, if $E_{u,1} - E_{u,2} \ll kT$. 
For example, \citet{J15b} used the LTE and optically thin formula for [Ni II] 7378 and [Fe II] 7155 to estimate the Ni/Fe mass ratio in core-collapse supernovae.



\subsubsection{LTE, optically thick case}
In the optically thick limit, $\beta_S \rightarrow 1/\tau_S$, and
\begin{equation}
L = 8 \pi \lambda^{-3} h\nu V t^{-1} \frac{ g_l n_u}{g_u n_l \left(1-\frac{g_l n_u}{g_u n_l}\right)}
\end{equation}
For LTE, $\frac{g_l n_u}{g_u n_l} = e^{-h\nu/kT}$, and so
\begin{eqnarray}
L = 8 \pi \lambda^{-3} h\nu V t^{-1} \frac{1}{e^{h\nu/kT}-1}\\
= 4 \pi V \left(ct\right)^{-1} \lambda \frac{2hc^2}{\lambda^5}\frac{1}{e^{h\nu/kT}-1}\\
=\frac{4 \pi V}{ct} \lambda B_\lambda(T)
\end{eqnarray}
The luminosity thus depends on the volume and the temperature, but not the mass.
Because the line width is proportional to $\lambda$, the
peak spectral flux is proportional to $B_\lambda(T)$. The peak flux values of
separated lines (and ignoring further transfer effects) would therefore follow a blackbody function.

We can use optically thick LTE lines to determine the volume of the emitting region if we know the temperature, or the temperature if we know the volume. 
Volume determinations are most robust if $h\nu \ll kT$, for which we get
\begin{equation}
L = 8 \pi \lambda^3 t^{-1} \mathcal{V} kT~~~~~~(h\nu \ll kT)
\end{equation}

Because the
volume \emph{span} can be directly inferred from the line expansion velocities ($\mathcal{V}_{span}=4\pi/3 \left(V_{max}t\right)^3$), in practice this means
that $\mathcal{V}/\mathcal{V}_{span}$ gives us the filling factor $f$ of the emitting region; what fraction of the volume is
effectively responsible for emission of that line. 

\citet{Li1993} performed LTE modelling for MIR lines with optical depth effects to estimate $f$ and $T(t)$ in SN 1987A. 
\citet{Jerkstrand2012} showed that [Fe II] 17.94 $\mu$m and [Fe II] 25.99 $\mu$m fall in the optically thick LTE regime for many hundred days in Type II models, and determined $f$ in SN 2004et. 
In Type Ia SNe, [Ni II] 6.634 $\mu$m, [Ni III] 7.350 $\mu$m, [Co III] 11.88 $\mu$m, and [Fe II] 17.93 $\mu$m can similarly be useful probes \citep{Maeda2010}.


From Table \ref{table:LTElines}, there are few optical/near infrared (NIR) lines where one can be confident to be in the optically thick LTE regime. [Ca II] 7291, 7323 is a candidate in SNe where the emission is from the synthesized calcium. In the early nebular phase, [O I] 6300, 6364, [Si I] 1.64 $\mu$m, [S I] 1.08 $\mu$m, and [Co II] 1.02 $\mu$m would also in many situations be in this regime. 

\subsection{Non-Local Thermodynamic Equilibrium}
\label{sec:NLTE}
\index{NLTE, nebular phase}
Outside LTE, the limiting formula depends on which mechanisms are assumed to dominate the populations and depopulations of the upper state. In general, one obtains expression involving the number abundance of the feeding state, and physical quantities such as temperature and electron density. If the goal is to estimate ion masses, it is desireable that this feeding state should be a ground state, as one may often approximate the mass of ions in the ground state to equal to the total element mass. Sometimes the coupling to a ground state can occur in several steps, as in recombination cascades.

The most common populating mechanisms are thermal collisions, non-thermal collisions, photoexcitation, and recombination. We will here consider two cases in particular, thermal collisionally excited lines, and recombination lines.

\subsubsection{Thermal collisionally excited lines}
The statistical equilibrium is, letting $f$ denote the feeding state (which may not be the same as the lower state $l$), and $N= n \mathcal{V}$
\begin{equation}
N_f Q_{uf}(T)\frac{g_u}{g_l}e^{-\left(E_u-E_f\right)/kT}n_e = N_u \left(A_{ul}\beta_{S,ul}  + Q_{ul}(T)n_e\right)
\end{equation}
Let us consider the regime $n_e \ll n_e^{crit}$. Then
\begin{equation}
N_u = N_f \frac{Q_{uf}(T)\frac{g_u}{g_f}e^{-\left(E_u-E_f\right)/kT}n_e}{A_{ul} \beta_{S,ul}}
\end{equation}
 The line luminosity becomes, using Eq. \ref{eq:basic}
\begin{equation}
L = \left(\mu m_p\right)^{-1}h\nu \times M_f Q_{uf}(T)\frac{g_u}{g_f}e^{-\left(E_u-E_f\right)/kT}  n_e  
\end{equation}
To determine the mass $M_f$ we would need to know both temperature and electron density. If neither is known, these lines have to be used in conjunction with other lines to make combined constraints. For example, if we have two emission lines from the same ion, being pumped from the same feeder state, then
\begin{equation}
\frac{L_1}{L_2} = \frac{h\nu_1}{h\nu_2}\frac{Q_{1f}(T)}{Q_{2f}(T)}  e^{-\left(E_2-E_1\right)/kT}
\end{equation}
Thus, such a line ratio may be used to determine the temperature at late times (when NLTE and optically thin conditions are more likely).

\subsubsection{Recombination lines}
\index{Recombination lines}

If the upper level is populated by recombinations (directly and/or through cascades via recombinations to higher levels), the equilibrium is (assuming $u$ is predominantly emptied in the $ul$ transition)
\begin{equation}
N_+ n_e \alpha_u^{eff}(T) = N_u A_{ul} \beta_{S,ul}
\end{equation}
where $\alpha_u^{eff}$ is the effective recombination rate. Then
\begin{equation}
L =  h\nu \left(\mu m_p\right)^{-1} \times M_+  n_e \alpha_u^{eff}(T)
\end{equation}
Determination of the mass $M_+$ of the recombining ion requires knowledge of the electron density and the temperature. The temperature-dependency of the effective recombination rates are, however, moderate, and so a determination of $M_+ n_e$ with relatively small uncertainty is possible. Recombination lines from two different elements residing in the same zone allows an estimate of the ratio of ion masses, because the electron density cancels out, and the ratio of effective recombination rates will often be almost temperature-independent. 

Emission lines from levels more than a few eV above the ground state are often powered by recombination.
Examples include H Balmer lines, some He lines, O I 7774, and Mg I 1.50 $\mu$m. Techniques to use O I 7774 as a diagnostic line are discussed in \citet{Maurer2010}, and the use of O I  7774, O I 9263, O I 1.129+1.130 $\mu$m, O I 1.316 $\mu$m and Mg I 1.50 $\mu$m recombination lines in \citet{J15a}.

\subsection{Discussion}
As we have seen, the luminosity in different line limits depends on different combinations of mass, volume, temperature, and electron density.
In general one needs multiple lines, preferably formed in different limits, to break the degeneracies and determine unique values for these parameters. 

Analytic line formation limits are easy to use, provide an understanding of line luminosity evolutions, and can be very useful diagnostics for some lines. The key to their application is to provide
convincing arguments for the validity of the physical regime. This in turn can come from three
different approaches: 1) Demonstrated good validity for any \emph{reasonable} physical conditions. 2) Inferrence from observations. 3) By inspection of forward models.

An added difficulty in SNe, compared to for example HII regions, is the high expansion velocities
which makes many lines blended with each other. Consideration of possible blending contaminations should always be done.

\newpage
\section{Radioactive powering} 
\index{Radioactive powering}
\label{sec:2}
\citet{Baade1945} discovered that supernovae decline on exponential tails with a time-scale of about 70d. The evolution of an explosion without further energy input produces neither enough luminosity nor such an exponential behaviour. This led \citet{Borst1950} to suggest that there is a radioactive power source. Following this were 20 years of speculation on what radioisotope this could be. Finally \citet{Colgate1969} provided the right answer; it is the second stage of the radioactive decay chain \ni56 $\rightarrow$ \co56 $\rightarrow$ $^{56}$Fe. \co56 decays on a time-scale of 111d to $^{56}$Fe, and at the same time the source of iron in the Universe had been identified.
The solution came after the demonstration of \ni56 production in high-temperature silicon burning by
\citet{Bodansky1968}.


In SNe, the relevant decay processes are electron captures (EC) and $\beta^+$ decays. 
In the first step the nucleus transmutes by the conversion of a proton to a neutron. In electron captures this energy is emitted as a neutrino, and in $\beta^+$ decays the energy is shared between a neutrino and the positron. The nucleus is usually left in some excited state, which then cascades to the ground state by emission of gamma-rays or by ejecting an inner-shell electron in an internal conversion. Following both electron captures and internal conversions, further emission of X-rays and/or Auger electrons occur as the inner hole is filled. The positrons annihilate with free electrons when they are slowed down to thermal energies, producing two 511 keV gamma rays for antiparallel spins and three gamma-rays with total energy 1022 keV for parallel spins.

The decay power is
\begin{equation}
P(t) = N(t)\frac{\hat{Q}}{\tau}
\label{eq:decay1}
\end{equation}
where $N(t)$ is the number of isotopes, $\hat{Q}$ is the average decay energy excluding neutrinos, and $\tau$ is the decay time scale. Note that $\hat{Q}$ is different from the normal $Q$-value, because a significant part of the decay energy will be emitted as neutrinos that escape the remnant.

For a primary isotope (like \ni56)
\begin{equation}
N(t) = N_0 e^{-t/\tau}
\end{equation}
whereas for a secondary isotope (like \co56)
\begin{equation}
N(t) = N_0 \frac{\exp{\left(-\frac{t}{\tau}\right)}- \exp{\left(-\frac{t}{\tau_p}\right)}}{1 - \frac{\tau_p}{\tau}}
\label{eq:decay2}
\end{equation}
where $\tau_p$ is the parent decay time scale (e.g. 8.8d for \ni56). We can write
\begin{equation}
N_0 = \frac{M_0}{\mu m_p} 
\label{eq:decay3}
\end{equation}
where $M_0$ is the mass of the primary species (weight $\mu$) synthesised in the explosion. 
Also other radionuclides are made in the explosion and may provide important input once most of the $^{56}$Co has decayed after a few hundred days. These include $^{57}$Ni/$^{57}$Co, $^{44}$Ti/$^{44}$Sc, $^{55}$Co/$^{55}$Fe, $^{22}$Na, and $^{60}$Co.
Decay data for the most common isotopes are listed in Table \ref{table:radionuclides}.


\begin{table}
\centering
\begin{tabular}{cccccccc}
\hline
Decay     & $\tau$ & $\hat{Q}$ (keV decay$^{-1}$) & $f_\gamma$ &  $f_{e+}$ &$f_{e-}$ & $f_{X-ray}$ & Source \\
\hline
$^{56}$Ni $\rightarrow$ $^{56}$Co & 8.77d & 1724 & 0.996 & 0 & 0.0025 & 0.0013 & Si burning\\
$^{56}$Co $\rightarrow$ $^{56}$Fe & 111d & 3732 & 0.966 (0.053) & 0.032 & 0.001 & 0\\
$^{57}$Ni $\rightarrow$ $^{57}$Co & 35h   &  2096& 0.924 (0.21) & 0.074 & 0.001 & 0 & Si burning \\
$^{57}$Co $\rightarrow$ $^{57}$Fe & 391d & 143  & 0.850 & 0 & 0.125 & 0.025 \\
$^{44}$Ti $\rightarrow$ $^{44}$Sc & 88y & 150 & 0 & 0 & 0.070 & 0.929 & $\alpha$-rich freeze-out\\
$^{44}$Sc $\rightarrow$ $^{44}$Ca & 6h & 2732 & 0.78(0.35) & 0.22 & 0 & 0\\
$^{55}$Co $\rightarrow$ $^{55}$Fe & 17h & 2429 & 0.822 & 0.18 &  0.001 & 0 & Si burning\\
$^{55}$Fe $\rightarrow$ $^{55}$Mn & 3.87y & 5.6 & 0 & 0 &  0.71 & 0.29 & \\

$^{22}$Na $\rightarrow$ $^{22}$Ne & 2.60y & 2392 & 0.918(0.38) & 0.082 & 0 & 0 & C burning \\
$^{60}$Co $\rightarrow$ $^{60}$Ni & 5.27y & 2600 & 0.963 & 0 &  0.037 & 0 & C burning\\ 
\hline
\end{tabular}
\caption{Decay data for the most common radioactive isotopes. Annihilation radiation from positrons is included in the gamma fraction (contribution in parenthesis).
Fractions smaller than $10^{-3}$ have been put to zero. Data from Nuclear Data Sheets 2011.}
\label{table:radionuclides}
\end{table}


%
%

\subsection{Deposition of decay products}
\label{subsec:2}

\subsubsection{Gamma-rays} 
\index{Gamma ray deposition}
The gamma-rays from radioactive decays are at MeV energies. 
The most important degradation process is Compton scattering. Because the gamma-ray energy is much higher than the excitation and ionization potentials of bound electrons ($\lesssim 1$ keV), the opacity is almost independent of the physical state of the matter - the cross sections
for incoherent scattering on various bound electrons are similar to those for free electrons, given by the Klein-Nishina formula. The local absorption thus depends only on the total density of electrons (free + bound).

For $E_\gamma > 1.022$ MeV pair-production can occur, and for $E_\gamma \le 0.1$ MeV photoelectric absorption becomes important as well. Photoelectric absorption will introduce some dependency on composition, but a relatively small fraction of the initial decay energy will be absorbed by this process.


When the gamma-ray Compton scatters, it loses some fraction of its energy and changes its direction. A detailed solution to the energy deposition requires computation of the multiple scattering processes. The first such calculation in the supernova context was carried out by \citet{Colgate1980}. The transfer process can be quite well described with
a gray opacity $\kappa_\gamma = 0.03 (Y_e/0.5)$ cm$^2$ g$^{-1}$ for the case of \co56 gamma-rays, where $Y_e$ is the ratio of the total number of electrons to the total number of nucleons. With this opacity, a uniform sphere becomes optically thin to the gamma-rays at
\begin{equation}
t_\gamma^{trap} = 33d \left(\frac{M}{1~M_\odot}\right) \left(\frac{E}{10^{51}~\mbox{erg}}\right)^{-1/2}
\end{equation}
For Type Ia and some Type Ib/c SNe gamma-ray escape needs to be considered already during the diffusion phase, whereas Type II SNe ($M\sim 10$) enter the tail phase (100-200d) still well before any significant escape occurs; this offers an opportunity to determine the \co56 mass by measuring the bolometric luminosity. Once gamma-ray escape has begun, fitting the observed bolometric light curve in the tail phase to a function such as $P(t)\times \left(1-e^{-t/\tau_\gamma}\right)$, can give an estimate for $\tau_\gamma$, and from that $ME^{-1/2}$.



\begin{figure}[htb]
\centering
\sidecaption
\includegraphics[width=0.8\linewidth]{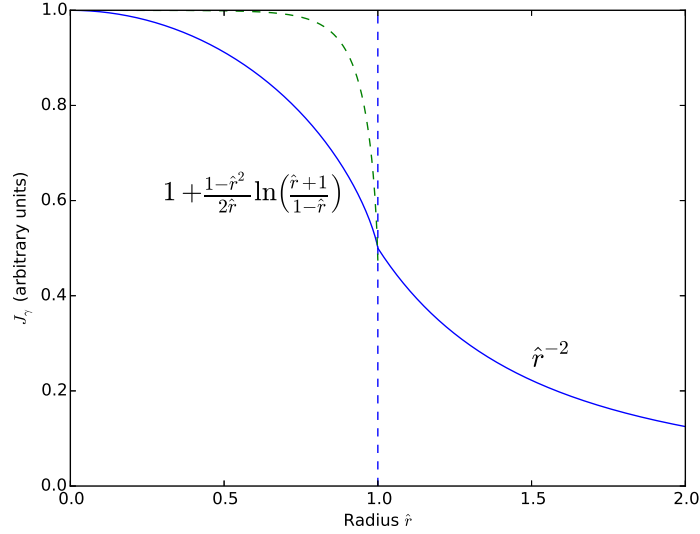} 
\caption{The solid line shows the gamma-ray radiation field for a uniform source distribution (between $0<\hat{r}<1$), in the optically thin limit. The dashed line shows the numerical solution for $\tau_\gamma=10$.}
\label{fig:gammadep}
\end{figure}

If both the radioactive source and ejecta are distributed as a uniform sphere, the mean intensity is, with $\hat{r}=r/R_{max}$
\begin{eqnarray}
J(\hat{r}) = \frac{1}{2} \int_{-1}^1 I(\hat{r},\mu) d\mu \\
= S_\gamma \frac{1}{2} \int_{-1}^{1}  \left(1 - e^{-\tau_\gamma \sqrt{1-\hat{r}^2(1-\mu^2)} - \tau_\gamma \hat{r}\mu}\right) d\mu
\end{eqnarray}
where $S_\gamma$ is the source function. This integral has no analytic solution, but for the $\tau_\gamma \ll 1$ limit
\begin{eqnarray}
J(\hat{r}) = \frac{1}{2}S_\gamma \tau_\gamma \int_{-1}^1 \sqrt{1-\hat{r}^2\left(1-\mu^2\right)} + \hat{r}\mu d\mu \\
= \frac{1}{2}j_\gamma R_{max} \left[1 +  \frac{1-\hat{r}^2}{2\hat{r}}\ln\left(\frac{\hat{r}+1}{1-\hat{r}}\right)\right]
\end{eqnarray}
where $j_\gamma$ is the gamma-ray emissivity. This function has its maximum at $\hat{r}=0$ and a value of half the maximum at $\hat{r}=1$ (see also \citet{Kozma1992}). For $\hat{r}>1$, the intensity falls off as $\hat{r}^{-2}$. Figure \ref{fig:gammadep} shows the $J(\hat{r})$ function in the optically thin limit, as well as the numerical solution for $\tau=10$. 

The deposition per mass is $4\pi J_\gamma \kappa_\gamma$. Thus, Fig. \ref{fig:gammadep} helps to envision how the absorption in a shell depends on its location with respect to the radioactive source (in the optically thin case).
Core-collapse explosion models predict density profiles that roughly follow $\rho \propto r^{-2}$ in the inner layers steepening to $r^{-10}$ in the outer. Thus, the bulk of the mass is relatively well described by a set of equal-width shells of similar mass ($4\pi r^2 \Delta r r^{-2}$ = constant). 


Because the gamma field intensity depends sensitively on location, it is important for modelling to use realistic ejecta models capturing the outcome of the mixing processes occurring in the explosion. For Type IIP
SNe the situation is relatively satisfactory, because simulations have shown that the metal regions
approach the limit of complete macroscopic mixing, and it should be a good approximation to have the \ni56, Si, and O zones occupy the same volume (macroscopically but not microscopically mixed). For stripped envelope SNe the situation is less clear and modelling becomes more uncertain. 

\subsubsection{Leptons}
\index{Positrons}
The leptons deposit their energy by colliding with the bound and free electrons in the SN ejecta. The effective opacity is around $\kappa_{e+} = $10 cm$^2$ g$^{-1}$ \citep{Colgate1980, Axelrod1980} for positrons, and similar for electrons.

Because the leptons are trapped for much longer than the gamma-rays, there is a phase where they take over as the dominant power source. This occurs when $\tau_\gamma \approx f_{lep}/f_\gamma\ (= 0.033\ \mbox{for}\ ^{56}\mbox{Co})$ (Table \ref{table:radionuclides}), which for a uniform sphere is at
\begin{equation}
t_{lep} = 180d \left(\frac{M}{1\ M_\odot}\right) \left(\frac{E}{10^{51}~\mbox{erg}}\right)^{-1/2} \left(\frac{f_{lep}/f_\gamma}{0.032}\right)^{-1/2}
\end{equation}
A successful identification of this transition phase would allow an estimate of $M E^{-1/2}$. Because $^{57}$Co takes over powering at $\sim$1000d, this transition never occurs in Type II SNe for $^{56}$Co positrons ($M \sim 10~M_\odot \rightarrow t_{lep} = 1800d$), but in Type Ia and Type Ib/c SNe the predicted transition occurs somewhere between 250d ($M=1.4~M_\odot,E=1$) and 900d ($M=5~M_\odot,E=1$). Unless other effects come into play, the bolometric light curve will then flatten onto the \co56 decay rate, having been steeper before due to $\gamma$-ray escape. In practice, it has proven difficult to demonstrate this transition for several reasons. SNe are dim at late times and estimating the photometry is difficult, with crowded fields and possible contamination by a remnant or companion. The temperatures are low and the true bolometric luminosity is difficult to determine lacking NIR and MIR observations (with molecule and dust formation adding to the problem). Finally, time-dependent effects (freeze-out) come into play at similar epochs and also lead to flattening of the light curve. 

The leptons may also escape eventually. If there is no magnetic field, or if the magnetic field is radially combed, their trapping time is about 20 times longer than the gamma-ray trapping time ($t_{lep}^{trap}=620d \left(M/1~M_\odot\right)\left(E/10^{51}~\mbox{erg}\right)^{-1/2}$). But a non-ordered magnetic field even at weak levels will lock the positrons in Larmor orbits and keep them trapped on very small scales:
\begin{equation}
\frac{R_L}{R_{max}} = 1.8\e{-6} \left(\frac{B}{10^{-6}~G}\right)^{-1} \left(\frac{V}{3000~\mbox{km s}^{-1}}\right)^{-1} \left(\frac{t}{100\mbox{d}}\right)^{-1}
\end{equation}
where $R_L$ is the Larmor radius.


\subsection{Degradation of non-thermal electrons}
\label{sec:nt}
\index{Non-thermal electrons}
After a few scatterings, the gamma-rays have been downgraded to $\lesssim 0.1$ MeV and are quickly photoabsorbed. In their wake we have a set of mildly relativistic electrons with typical energies between 0.1-1 MeV. These electrons in turn lose their energy by collisions with free and bound electrons leading to ionization, excitation, and heating of the gas. This process is familiar to us from a closer-to-home environment of Aurorae. The collisional ionizations lead to the creation of further high-energy electrons, generally referred to as secondaries. The secondaries have, however, much lower energies than the primaries, $\sim 10-100$ eV.

The degradation process can be modelled with Monte Carlo methods \citep{Shull1979, Fransson1989}, or formal solutions \citep{Lucy1991, Xu1991, Kozma1992}. A simplified formal solution is the continuous-slowing-down approximation \citep{Axelrod1980}, which is quite accurate for the primary electrons, but not for the secondaries.

An important property of this process is that the solution has no strong dependence on the energy of incoming primary particles, as long as its over $\sim$ 1 keV. This allows
for generic solvers that do not need information about which high-energy particle source is involved.
There is also only a weak dependency on density, leaving the relative abundances of ions and electrons as the main parameters of the problem.

\textbf{Ionization.} Non-thermal ionization determines the ionization balance for the major species and thereby
which ions form the spectrum. To solve for the fractions of energy going into different channels, one needs to know the differential cross sections for the collisional ionization processes, i. e. $\sigma(E_{in}, E_{out})$. For many elements these differential cross sections are not known, neither experimentally nor observationally, and an approximate treatment is needed. The method of \citet{Kozma1992}, for example, applies a differential form measured for O I to all ions.

For the integrated cross sections, for electron energies much higher than the excitation/ionization potential $I$, one may use the Bethe approximation for both ionization and excitation cross sections. The cross section for ion $i$, transition $j$, is
\begin{equation}
\sigma_{coll,ij}(E)= \frac{\pi a_0^2}{E}\left(c_{1,ij}\ln E +c_{2,ij}\right)~~~E \gg I~,
\end{equation}
where $E$ is the energy of the electron, $a_0$ is the Bohr radius, 
and $c_{1,ij}$ and $c_{2,ij}$ are constants depending only on level energies and transition strengths. For lower energies, specific calculations or experimental data for the collision cross sections are needed. 


For even moderately ionized plasmas with $x_e \gtrsim 0.1$, only a moderate fraction, $\sim 1-20\%$ of the energy goes to ionization. This increases towards a plateau value of $\sim$50\% for more neutral gas.

\textbf{Excitation.} The coefficient $c_{1,ij}$ is, for bound-bound transitions, proportional to the oscillator strength $f_{ij}$. Allowed transitions ($f_{ij}\sim 1$) will therefore in general be more important than forbidden ones ($f_{ij} \ll 1$). 
Non-thermal excitations tend to populate high-lying states connected to the ground state by allowed transitions, which leads to UV emissivity. Because of both photoelectric and line opacity, this UV emission does not escape directly, but scatters and undergoes fluorescence. For this reason, models that include non-thermal excitation should preferably also include radiative transfer.

Consideration of non-thermal excitation becomes more important for later epochs; the declining electron fraction gives less heating and more energy going into excitation (and ionization), and the declining temperature makes the resulting fluorescence into the optical and NIR more prominent as thermal emission shifts into the MIR (the so-called ``infrared catastrophe'').
However, these processes also affect the spectrum in indirect ways. The increased population of excited states can lead to photoionization from these levels and a higher electron fraction.
For example, H$\alpha$ in Type II SNe arises mainly as a non-thermal excitation first populates $n=2$, the H I atom is then photoionized from this state, and in the ensuing recombination process H$\alpha$ is emitted (see also Sec. \ref{sec:Hlines}).

\textbf{Heating.} As long as the plasma is ionized to $x_e \gtrsim 0.1$, most of the non-thermal energy goes to heating of the gas. The fraction grows monotonically with increasing ionization state. The SN spectrum will therefore be dominated by cooling emission, which is dominated by lines. In Type Ia SNe, the ionization state is high compared to core-collapse SNe, and \citet{Axelrod1980} obtained solutions of the heating fraction close to unity. 
The ionization balance depends of course critically on the exact value of the ionization fraction, even if it is small. 

\begin{figure}[htb]
\centering
\includegraphics[width=0.49\linewidth]{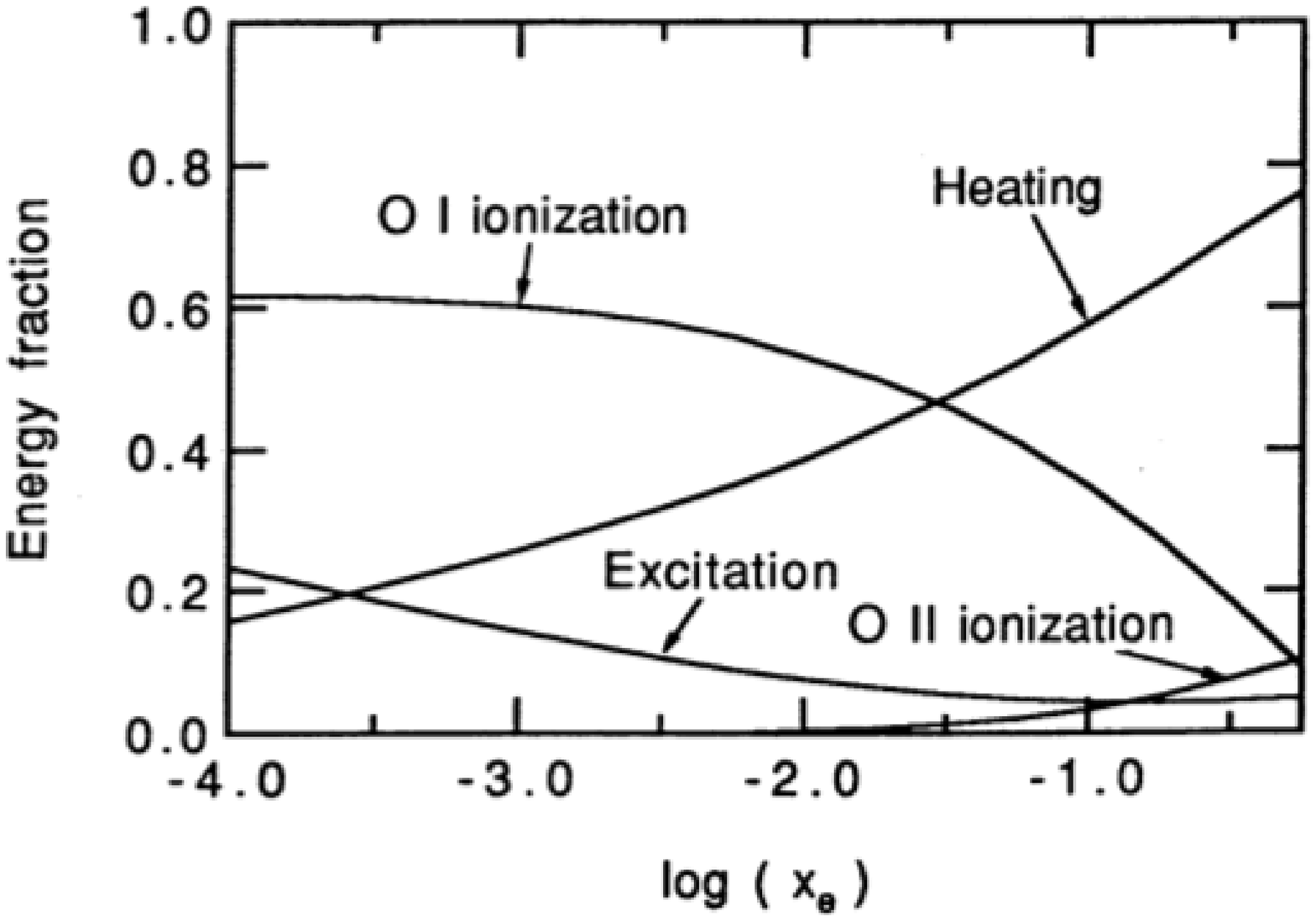}
\includegraphics[width=0.49\linewidth]{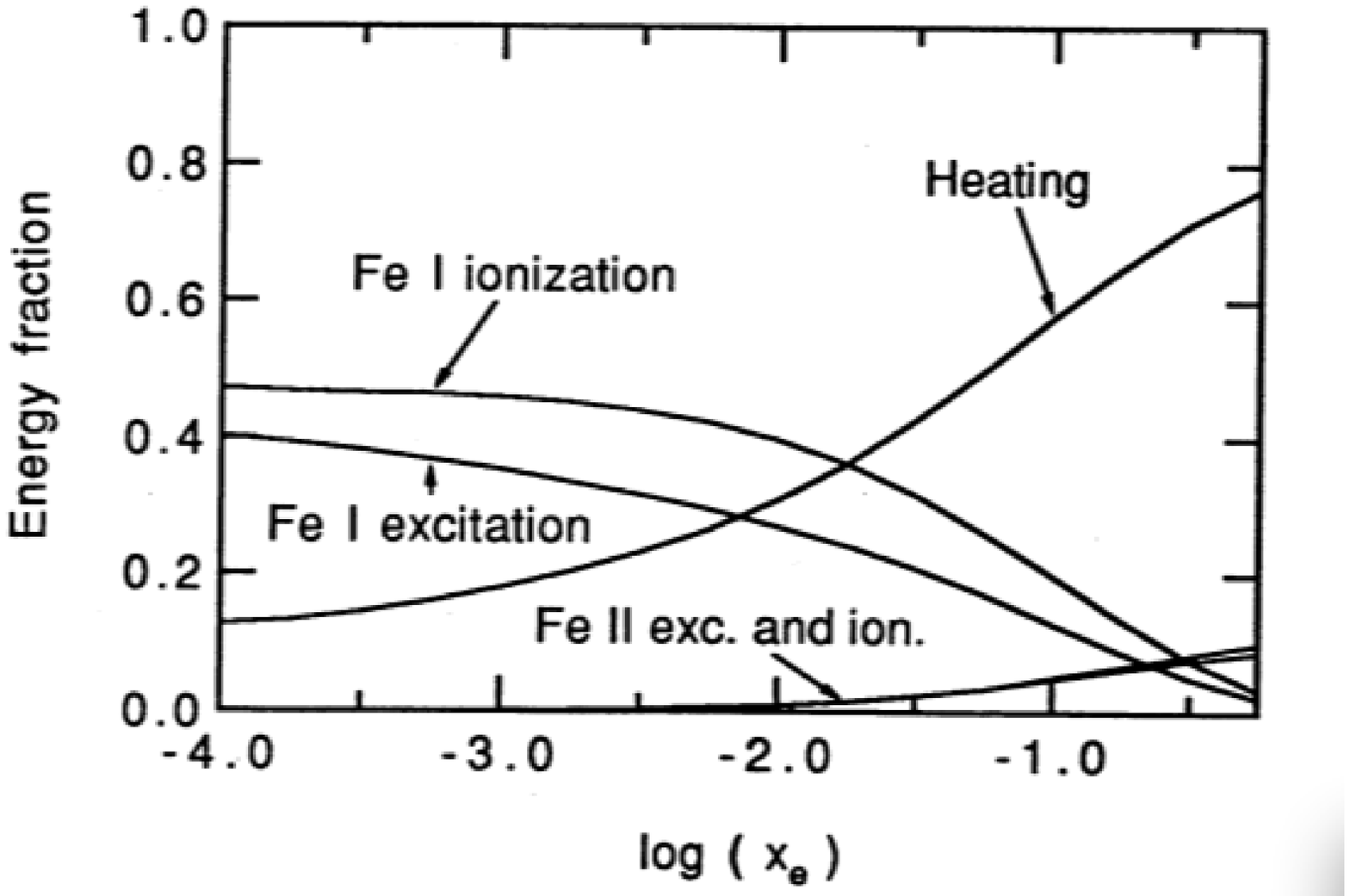}
\caption{The fraction of radioactive powering going into heating, ionization, and excitation in a pure oxygen plasma (left) and a pure iron plasma (right), as function of the electron fraction $x_e$. From \citet{Kozma1992}.}
\label{fig:kf92}
\end{figure}

Figure \ref{fig:kf92} shows example solutions for pure oxygen and iron plasmas in the low-ionization limit (only neutral and first ionization stages present). The distribution of non-thermal electrons, ionization, temperature, and excitation are all connected equation systems. Thus, one needs to iterate - the non-thermal rates are calculated given an ionization and excitation structure. These are then updated (also following updates of temperature, radiative rates).


\newpage
\section{Spectral modelling} 
\index{Spectral modelling, nebular phase}
\label{sec:modelling}

A spectral model defines a scope and a set of physical approximations and assumptions to compute the emergent flux. Spectral models can be divided into two categories 1) Models for individual line luminosities (or sets of lines) and 2) Models for the full spectrum of the SN (or some range of the spectrum). The modeller should ideally not only set up and 
compute the model, but also use physical reasoning to assess which predictions
are robust and which depend more sensitively on ill-constrained assumptions.


There is a large number of physical processes at play in supernova ejecta that each can be treated in several different approximations. 
The basic ingredients in a nebular model, with different levels of approximation, are listed below together with examples of models implementing them (for a series of papers by the same first-author using the same technique, only the first paper is listed).


\begin{itemize}
\item Nebula structure
\begin{itemize}
\item 1D, 1-zone \citep{Axelrod1980, RuizLapuente1992, Mazzali2001}
\item 1D, multi-zone, no mixing \citep{Fransson1989, RuizLapuente1995, Liu1997a, Sollerman2000, Mazzali2007}
\item 1D, multi-zone, artificial mixing \citep{Houck1996, Kozma1998a, Sollerman2000, Dessart2011, Jerkstrand2011} 
\item 2D \citep{Maeda2006a} 
\item 3D \citep{Kozma2005} 
\end{itemize}
\item Gamma-ray transfer 
\begin{itemize}
\item Local deposition \citep{Dessart2011} 
\item Gray transfer \citep{Axelrod1980, Houck1996, Liu1997a, Kozma1998a,  Mazzali2001, Jerkstrand2011} 
\item Compton scattering \citep{Fransson1989, Eastman1993}
\item Compton scattering + photoelectric \citep{Dessart2013}
\item Compton scattering + photoelectric + pair-production \citep{Maurer2011}
\end{itemize}
\item Non-thermal processes
\begin{itemize}
\item Pure heating \citep{Dessart2011} 
\item Fixed heating and ionization (0.97-0.03) \citep{RuizLapuente1992, Mazzali2001}
\item Heating, ionization, excitation - continuous slowing down \citep{Axelrod1980,Eastman1993}
\item Heating, ionization, excitation - full solution\citep{Kozma1992, Liu1997a, Jerkstrand2011, Dessart2013}
\end{itemize}
\item Level populations
\begin{itemize}
\item LTE
\item NLTE with collisional excitation/deexcitation and radiative decay terms \citep{Axelrod1980,RuizLapuente1992, Mazzali2001}
\item NLTE with comprehensive set of processes \citep{Fransson1989, Houck1996, Maurer2011, Jerkstrand2012}
\item NLTE with comprehensive set of processes, time-dependent \citep{Kozma1998a, Dessart2013}

\end{itemize}
\item Temperature 
\begin{itemize}
\item Non-thermal heating, line cooling \citep{Axelrod1980,RuizLapuente1992, Mazzali2001}
\item Non-thermal and photoelectric heating, line, recombination, and free-free cooling \citep{Houck1996, Jerkstrand2011, Maurer2011}
\item Non-thermal and photoelectric heating, line, recombination, and free-free cooling, time-dependent \citep{Kozma1998a}
\end{itemize}
\item Radiative transfer
\begin{itemize}
\item Sobolev approximation locally, no global transport \citep{Liu1997a, Kozma1998a, Mazzali2001, Maeda2006a} 
\item Sobolev approximation locally, global transport \citep{Jerkstrand2011, Maurer2011}
\item Full transport locally and globally, time-dependent \citep{Dessart2011} 
\end{itemize}

\end{itemize}

The number of atoms and levels modelled, and the quality of the atomic data library have large impact on the accuracy of the model. Before comparing a model to data, it is important to understand the set-up and limitations, and assess the ability to
predict any given observable.

Gamma-ray deposition and non-thermal processes have been covered already, and we discuss the other components in more detail here.


\subsection{Nebula structure} 
\label{sec:mixing}

One of the biggest challenges to SN spectral modelling is to capture the complex structures evidenced by observations of SN remnants, and also obtained in multi-D explosion simulations. Most models are set up in 1D, and some consideration of these mixing effects is needed. It is important to distinguish between \emph{microscopic} and \emph{macroscopic} mixing.

\textbf{Microscopic mixing.} 
\index{Mixing}
Each nuclear burning stage gives unique nucleosynthesis products, and these do not readily become mixed with each other on atomic scales in the SN ejecta because the supersonic flow freezes out composition on velocity scales $v_{therm}(t)$, which is $\lesssim$ 10 \kms already after a few days. Diffusion is also inefficient in the very early phases when temperatures are still high. 
The cross section for atomic collisions is of order $\sigma=\pi a_0^2 \sim 10^{-16}$ cm$^2$, where $a_0$ is the Bohr radius. For ion collisions Coulomb interactions give somewhat larger values for typical temperatures.
The ``optical depth'' is $\tau=\sigma n L$ = $\sigma n Vt$. Of order $\tau^2$ scatterings are needed to travel a distance $\tau$. The time between scatterings is $\Delta t = \lambda/v_{therm} = 1/\sigma n v_{therm}$. The total time to diffuse through the nebula is
\begin{equation}
t_{atomic-diff} = \frac{\left(\sigma n L\right)^2}{\sigma n v_{therm}} = \frac{\sigma n L^2}{v_{therm}} = \frac{3 \sigma M}{4\pi V_{max}t v_{therm}(T) \bar{m}} 
\end{equation} 
For any reasonable values of $M,V_{max},T$, the diffusion time is longer than the age of the Universe, and neglegible mixing occurs. 

This idea is supported by several lines of evidence. For instance, models for molecule formation show that microscopically mixed ejecta fail to produce the observed amounts of molecules \citep[e.g.][]{Liu1996, Gearhart1999}. Spectral models with full microscopic mixing also generally fare worse than models without such mixing \citep{Fransson1989}.

\textbf{Macroscopic mixing.} Macroscopic mixing, on the other hand, is known to occur vigorously during the explosion of many progenitor structures. The dominant mechanism is typically Rayleigh-Taylor mixing, which arises behind the reverse shocks created at composition interfaces where $\rho r^3$ shows a positive derivative. The mixing effects are strong in explosion simulations of H-rich SNe, with red supergiants becoming more mixed than blue supergiants \citep[e.g.][]{Herant1994}. Simulations of He core explosions show strong mixing for low-mass progenitors, but weaker for high-mass ones \citep[e.g.][]{Shigeyama1990}.
%
The lack of O/He and He/H interfaces in Type Ic SNe makes mixing harder to obtain. 

The implication is that hydrogen-rich SNe are poorly represented by 1D explosion models, and either 2D/3D models or 1D models with artificial macroscopic mixing are preferred. For stripped-envelope SNe there is more freedom within current uncertainties in the mixing processes. In Type Ia SNe, Rayleigh-Taylor instabilities also occur, with buoyant hot $^{56}$Ni rising up between downflows of C and O, in particular in deflagration models \citep[e.g.][]{Roepke2007}. 




\subsection{Level populations}
\index{NLTE}
\index{Level populations, NLTE}

At nebular times LTE can in general not be assumed. The level populations have to be determined by solving a set of NLTE equations describing the various populating and depopulating mechanisms. Letting $n_{ijk}$ denote the number of ions of element $i$ ionization state $j$ and excitation state $k$, each equation has the form
\begin{equation}
\frac{dn_{ijk}}{dt} - \frac{n_{ijk}}{\rho}\frac{d\rho}{dt} = \sum_{i'} \sum_{j'} \sum_{k'} n_{i'j'k'} \mathcal{R}_{i'j'k',ijk} - n_{ijk} \sum_{i'} \sum_{j'} \sum_{k'}  \mathcal{R}_{ijk,i'j'k'}
\label{eq:sse}
\end{equation}
The rates $\mathcal{R}$ depend in general on $n_e$, $T$, radiation field $J_\nu$, and non-thermal electron distribution $n_{nt}(E)$. In the Sobolev approximation, they also depend on the level populations $n_{ijk}$ themselves. 
Charge transfer (ct) also implies dependencies on number densities of colliding partners. All normal processes have $i' = i$, but radioactive decay and molecule and dust formation allows $i' \neq i$. If we ignore this for now, so dropping the $i$ index, we can break up all processes populating/depopulating a level into the groups of ionization, recombination, excitation, and deexcitation.
Letting $R$ denote a radiative process, and $C$ a collisional, these group can be further broken up into
\begin{eqnarray}
\nonumber \mathcal{R}_{ion} &=& R_{photo-ion.} + C_{thermal\ ion.} + C_{non-thermal~ion.} + C_{ct,ion.}\\
\nonumber \mathcal{R}_{rec} &=& R_{stim.\ rec.} + R_{rad.\ rec.}  + R_{diel.\ rec.} + C_{3-body\ rec.} + C_{ct,rec}\\
\nonumber \mathcal{R}_{exc} &=& R_{photo-exc.} + C_{thermal~exc.} + C_{non-thermal~exc.}\\
\nonumber \mathcal{R}_{de-exc.} &=& R_{stim.\ em.} + R_{spont.~decay}  + C_{thermal~de-exc.} \\
\label{eq:apa}
\end{eqnarray} 
which gives an overview of the physical processes involved.
Let $N$ be the typical number of excited states modelled in each ion. Any given level can transition to any other
level in that ion, in the ion below, or above, i.e. to $\sim 3N$ other levels.
The number of transition pairs is $N_{ion-stages} 3N^2/2 \sim 10^4N_{ion-stages}$ per chemical species for $N=100$.
The modelling of just a few species would require specification of $\sim 10^5$ transition rates $\mathcal{R}$, each in turn containing 15 individual rates (Eq. \ref{eq:apa}), some of which are temperature dependent. The only rates that are available in batch reading on this scale are spontaneous radiative decay rates. Photoionization and non-thermal ionization rates requires cross sections as function of energy, and simplifying treatments, or inclusion of only some transitions, is necessary. The following considerations help:
\begin{itemize}
\item For excited non-metastable states, $\mathcal{R}_{exc} = \mathcal{R}_{ion}= \mathcal{R}_{rec}=0$, and $\mathcal{R}_{de-exc} = R_{spont.\ decay}$ are accurate approximations. These states, with $A\gg 1$, are empitied on fractions of a second ($\tau \sim 1/A$) by spontaneous decay, much shorter than any other process can operate.
\item For ground states, $C_{thermal.\ ion}=0$ and $C_{3-body\ rec.}=0$ are accurate approximations because nebular temperatures are too low for these processes to be competitive.
\end{itemize}


\index{Steady-state}
\emph{Steady-state} corresponds to ignoring the time-derivative terms in Eq. \ref{eq:sse}, so a (non-linear) algebraic equation system follows, which is straightforwardly solved by Newton-Raphson iteration. This is justified as long as the reaction time-scales are short compared to dynamic and radioactive time-scales. The slowest reaction is typically radiative recombination, where $t_{rec} \sim 1/\left(\alpha n_e\right) \sim 1d\left(10^7 \mbox{cm}^{-3}/n_e\right)$. 


The simplest possible scheme for NLTE excitation solutions is to include spontaneous radiative rates, treated in the Sobolev approximation, and
thermal collisional rates. This gives a reasonable approximation for plasmas with $x_e \gtrsim 0.1$ because heating then accounts for most of the energy deposition (Sect. \ref{sec:nt}), and the reprocessing of this energy is mainly done by collisional cooling.
The next natural step is to add recombination, which enables
accounting of a non-thermal ionization energy and production of recombination lines.
Coupling the radiation field to the NLTE solutions requires addition of line absorption and
photoionization rates.
The most advanced models include also non-thermal excitations, photoionization from excited states, and charge transfer reactions.



\index{Charge transfer}
Charge transfer (CT) is an important process at high densities and low ionization; a relatively unique astrophysical environment found in supernovae at late times. By this process electrons jump from one ion to another, for example
\begin{equation}
\mbox{O I + H II} \leftrightarrow \mbox{O II + H I}
\end{equation}
Both species in the outgoing channel may be left in excited states.
The important role of this process in governing the ionization balance in SNe was pointed out by \citet{Meyerott1978}. The fast rates are of order $\Gamma_{CT} \sim 10^{-9}$ cm$^3$ s$^{-1}$, to be compared with radiative recombination rates of order $\Gamma_{RR} \sim 10^{-12}$ cm$^3$s$^{-1}$. Thus, even an atomic abundance of $10^{-3}$ per electron can lead to recombination by charge transfer dominating the ionization balance.

The process is typically fast between neutral atoms and singly ionized ions, and if the energy defect is small. Ion-ion reactions require high temperatures due to the Coulomb barrier, and are of less importance. Modelling is hampered by many unknown or poorly known rates. Most rates that have been calculated involve H and He, for application in HII regions and planetary nebulae, but SN modelling mainly needs rates between metals.

When the reaction is fast in both directions, the effect of CT is to link the ionization balance of the less abundant element to the more abundant. As an example, for a solar composition gas the primordial O attains the same ionization balance as H by this mechanism.


\subsection{Temperature}
\index{Temperature, nebular phase}
The temperature evolution is obtained by solving the first law of thermodynamics, which for homologous expansion and pressure from a perfect monoatomic gas is
\begin{equation}
\frac{dT(t)}{dt} = \frac{H(T) - C(T)}{\frac{3}{2} k n(t)} - \frac{2T(t)}{t} - \frac{T(t)}{1+x_e(t)}\frac{dx_e(t)}{dt}
\label{eq:temperature}
\end{equation}
where $H(t)$ and $C(t)$ are the heating and cooling rates (which depend on the NLTE solutions $n_{ijk}$). The heating is usually dominated by non-thermal heating, and the cooling by collisional excitation of fine-structure lines (which then decay radiatively). 

The \emph{thermal equilibrium approximation} corresponds to setting $H(T) = C(T)$ and solving the resulting algebraic system for $T$; this is a good approximation if the cooling/heating time-scales are short compared to the dynamic and radioactive decay time scales. This is usually fine for several hundred days into the nebular phase. For example, \citet{deKool1998} and \citet{Kozma1998a} find this approximation to hold for 600-800d for the H zone and several years or decades for the metal zones in models for SN 1987A.
Once collisional cooling becomes slow, adiabatic cooling takes over, and the solution to Eq. \ref{eq:temperature} is $T(t) \propto t^{-2}$ (ignoring the last term involving $x_e$ which typically has a small effect).

Once molecules begin to form, the cooling becomes more efficient and the temperature can become significantly lower than models without molecules would suggest. While a few studies have adressed formation and cooling of molecules in single-zone setups, they are typically not included in multi-zone models.

\subsection{Radiative transfer}
\index{Radiative transfer, nebular phase}
   
\begin{figure}[htb]
\centering
\includegraphics[width=0.8\linewidth]{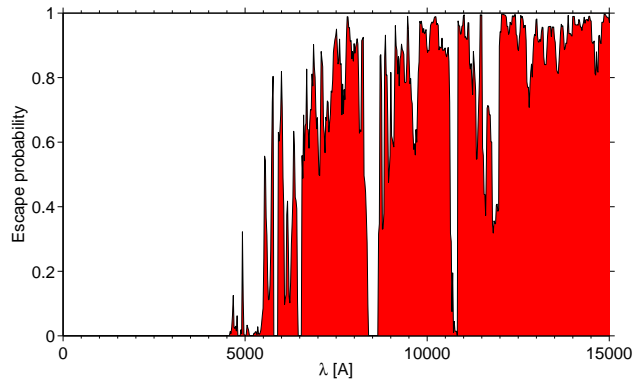}
\caption{The probability for a photon emitted from the center of the SN to escape the ejecta without being absorbed in a line, for a Type II model at 300 days. From \citet{J11thesis}.}
\label{fig:eprob}
\end{figure}

   
Supernova ejecta remain optically thick to line blocking below $\sim$ 4000-5000 \AA~for years or even decades. Fig. \ref{fig:eprob} illustrates this, showing the photon escape probability in a Type IIP SN at 300d.
Thus, to model the appearance of the SN at short wavelengths, as well as lines at longer wavelengths
influenced by fluorescence, this transfer must be considered.
There are two conceptual approaches - solution of the radiative transfer equation and Monte Carlo simulations of the photon propagation.

The transfer equation can be solved by the method of characteristics.
A common approach is to solve high-resolution moment equations (which remove
the angular variable) with a variable Eddington factor,
which is determined iteratively by a formal solution to the transfer equation \citep[e.g.][]{Hillier2012}.

\index{Monte Carlo}
The basic concept in the Monte Carlo approach is to follow energy packets as they
propagate through the ejecta, using random numbers to determine their interactions and trajectory changes.
\citep[e.g.][]{Jerkstrand2011}.
Two properties of the SN make the machinery relatively straightforward to set up -
homology (Hubble flow) makes the expansion isotropic from any point in the ejecta,
and the large velocity gradient allows the Sobolev approximation which means that
the photon can be transported from line to line, without having to consider line overlap.
Different algorithms may be chosen depending on the desired degree of coupling
to the gas state. 

Iteration between NLTE solutions and radiative transfer have good convergence properties in late-time SN environments, because the dominant excitation and ionization processes are typically collisional (non-thermal and thermal).
In practise, only lines and photoionization continua from ground states and (effectively) meta-stable states need to be considered. 

What happens to a photon absorbed in a resonance line?
For allowed lines, the Regemorter formula gives an estimate of the effective collision strength, $\Upsilon \sim 2\e{6}\lambda^3 A g_u P_y$, where for purposes here we can take $P_y = 0.1$. Then $Q_{res} \sim 5\e{-16} A \left(\lambda/3000~\AA\right)^3$ cm$^3$ s$^{-1}$.
The ratio of probabilities for collisional deexcitation (i.e. thermalization) and scattering is then
\begin{equation}
\frac{p_{therm}}{p_{scatter}} = \frac{Q_{res}n_e}{A/\tau_S} \approx \left(\frac{n_e}{10^8~\mbox{cm}^{-3}}\right)
\left(\frac{\tau_S}{10^7}\right) \left(\frac{\lambda}{3000~\AA}\right)^3
\end{equation}
Because $n_e \lesssim 10^8$ cm$^{-3}$ in the nebular phase 
only the most optically thick lines ($\tau_S \gtrsim 10^7$) may suffer thermalization.
Note that even if collisional deexcitation is unimportant, multiple scatterings may still lead to enhanced thermalization by other processes such as photoionization as the photons increase their dwell time in the ejecta. Fluorescence gives a capping of this trapping by moving photons out to optically thinner wavelengths after a handful of scatterings.


Modern codes treat the complex global transport through hundreds of thousands of lines. However, simplified approaches are also conceivable.
\citet{Axelrod1980} devised an approximate method to treat the global line transfer in a single-zone model. In this method, a position and angle averaged probability to be absorbed in any other transition (using the Sobolev approximation) is calculated (compare with the average continuum escape probability of \citet{Osterbrock1989}). To avoid coupling between the radiation field and the optical depths (the method is in the ``no coupling'' regime), it is then assumed that absorption occurs only from low-lying states whose populations are calculated without any coupling to the radiation field. The fluorescence is finally obtained by coupling the high-lying state populations through the averaged probabilities, which gives a linear system of equations. The method showed relatively good results for reproducing the blue regions of Type Ia SNe. While the method is outdated for modelling of quite well understood SN classes, it can still find use in rapid exploration of model scenarios for new SN types.

\newpage
\section{Application of spectral models}
\label{sec:examples}



Spectral models may be used to compare with observations to find the best matches in terms of progenitors, nucleosynthesis, and mixing. Overall, the application of models may be divided into:


\begin{itemize}
\item Test the viability of a particular explosion model for an observed SN or SN class
\item Estimate a physical parameter by optimising a model over this parameter
\item Identify lines
\item Determine physical conditions and regimes
\item Estimate bolometic correction factors
\end{itemize}

Apart from these concrete purposes, a model gives opportunity to improve our general understanding of what is going on in the SN. As in all science, the end use of such fundamental information cannot always be predicted. 

\subsection{Hydrogen-rich SNe}
\index{Type IIP SNe}
\index{Nebular models}
Nebular multi-zone spectral models of H-rich SNe have been presented by \citet{Fransson1987, Kozma1998a,Kozma1998b,deKool1998, Dessart2011,Dessart2013,Jerkstrand2011,Jerkstrand2012,Jerkstrand2014,J15b}. This 'first principles' model set is complemented by a set of more parameterized models studying specific line formation in H \citep{Xu1992}, He \citep{Li1995}, O \citep{Li1992}, Ca \citep{Li1993}, and Fe \citep{Li1993iron} in SN 1987A.

\begin{figure}[htb]
\centering
\includegraphics[width=1\linewidth]{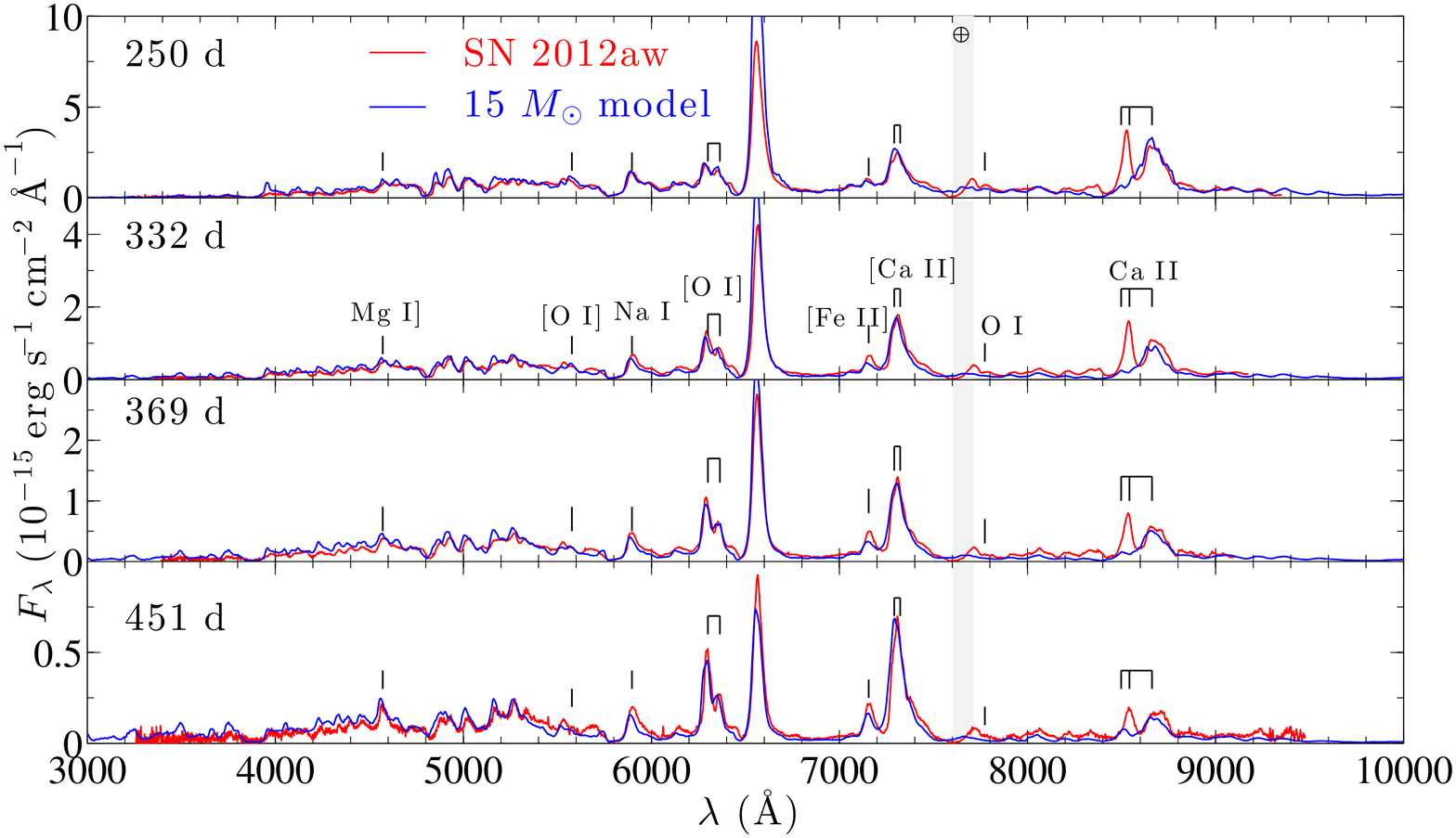}
\includegraphics[width=0.7\linewidth]{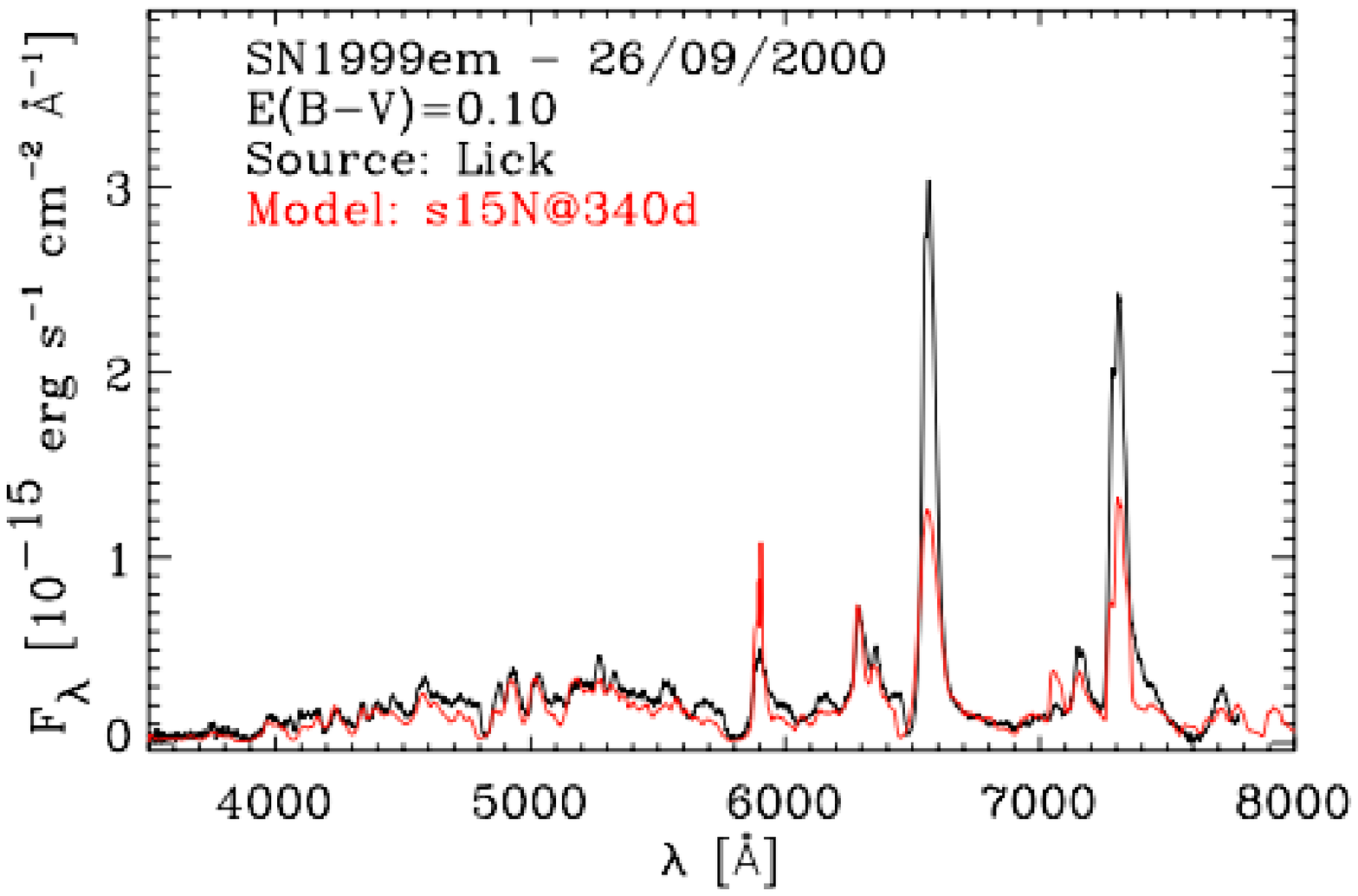}
\caption{Examples of Type IIP spectral models. \emph{Top:} A $M_{ZAMS}=15$ \msun~spectral model from \citet{Jerkstrand2014}, compared to SN 2012aw. \emph{Bottom:} A spectral model from \citet{Dessart2013}, compared to SN 1999em.}
\label{fig:IIPmodel}
\end{figure}

The output of the \citet{Fransson1987, Kozma1998a,Kozma1998b} and \citet{deKool1998} models (all for SN 1987A) are line luminosity tracks, whereas
the other papers present spectra. Figure \ref{fig:IIPmodel} shows two model examples, from \citet{Dessart2013} and \citet{Jerkstrand2014}. Current-day models are quite successful at reproducing the main spectral features, such as Mg I] 4571, Na I, [O I] 6300, 6364, H$\alpha$, [Fe II] 7155, [Ca II] 7291, 7323, and Ca II NIR, as well as the underlying quasi-continuum. Three of the most prominent lines in nebular Type IIP spectra are H$\alpha$, [O I] 6300, 6364, and [Ca II] 7291, 7323.
The formation of these lines is reviewed in some more detail below. 


\subsubsection{Hydrogen lines}
\index{Hydrogen lines, nebular phase}
\label{sec:Hlines}
The models show that the H lines are formed mainly by recombination.
The ionization is achieved in a two-step process; first $n=2$ is populated by
non-thermal excitations, and then follows a Balmer photoionization. The ionizing photons come partly
from H itself in the form of two-photon emission (from $n=2$), and partly from line emission from
other elements.


Apart from H$\alpha$, there are no other Balmer lines produced for several hundred days. \citet{Kirshner1975}, and later \citet{Xu1992} and \citet{Kozma1992}, demonstrated that this is because the Balmer series is optically thick (``Case C''), and conversion occurs through other series (e.g. H$\beta$ ($n=4$ to $n=2$) converts to Pa$\alpha$ ($n=4$ to $n=3$) + H$\alpha$ ($n=3$ to $n=2$). The Balmer lines are thick because $n=2$ acts like a meta-stable state due to the enormous optical depth in Ly$\alpha$ ($\tau_S \sim 10^{10}$). 


Modelling of the H lines is also complicated by breakdown of the Sobolev approximation
for the Lyman lines - at very high $\tau_S$ values one must consider line overlap as well as the possibility that the photons scatter into regions of the ejecta with different composition.
Considering Ly$\alpha$ escape, the $n=2$ population may decrease, which
in turn decreases the number of photoionizations and the recombination line
luminosities throughout. Considering Ly$\beta$ escape, the H$\alpha$ luminosity 
specifically may decrease while not impacting the other H lines. 

%
%
%
An important result from H line modelling in SN 1987A is that H-gas must
occupy most of the volume of the central few 1000 \kms, and absorb
about half of the gamma rays \citep{Xu1992}. Combined with the lack of flat-topped
line profiles, the picture is clear that strong mixing occurs in the
explosion and drags hydrogen envelope material down towards low velocities, in line with
multi-D hydrodynamic simulations.
\citet{Kozma1998b} estimate a total H-zone mass of 8 $M_\odot$ in SN 1987A (about half
of which is H, the rest He). Standard stellar evolution models with a $\sim$8 $M_\odot$ H-zone
give satisfactory fits to H-lines also in many other Type II SNe \citep{Jerkstrand2012,Jerkstrand2014}.


\subsubsection{Oxygen lines}
\index{Oxygen lines, nebular phase}

\begin{figure}[htb]
\centering
\includegraphics[width=0.8\linewidth]{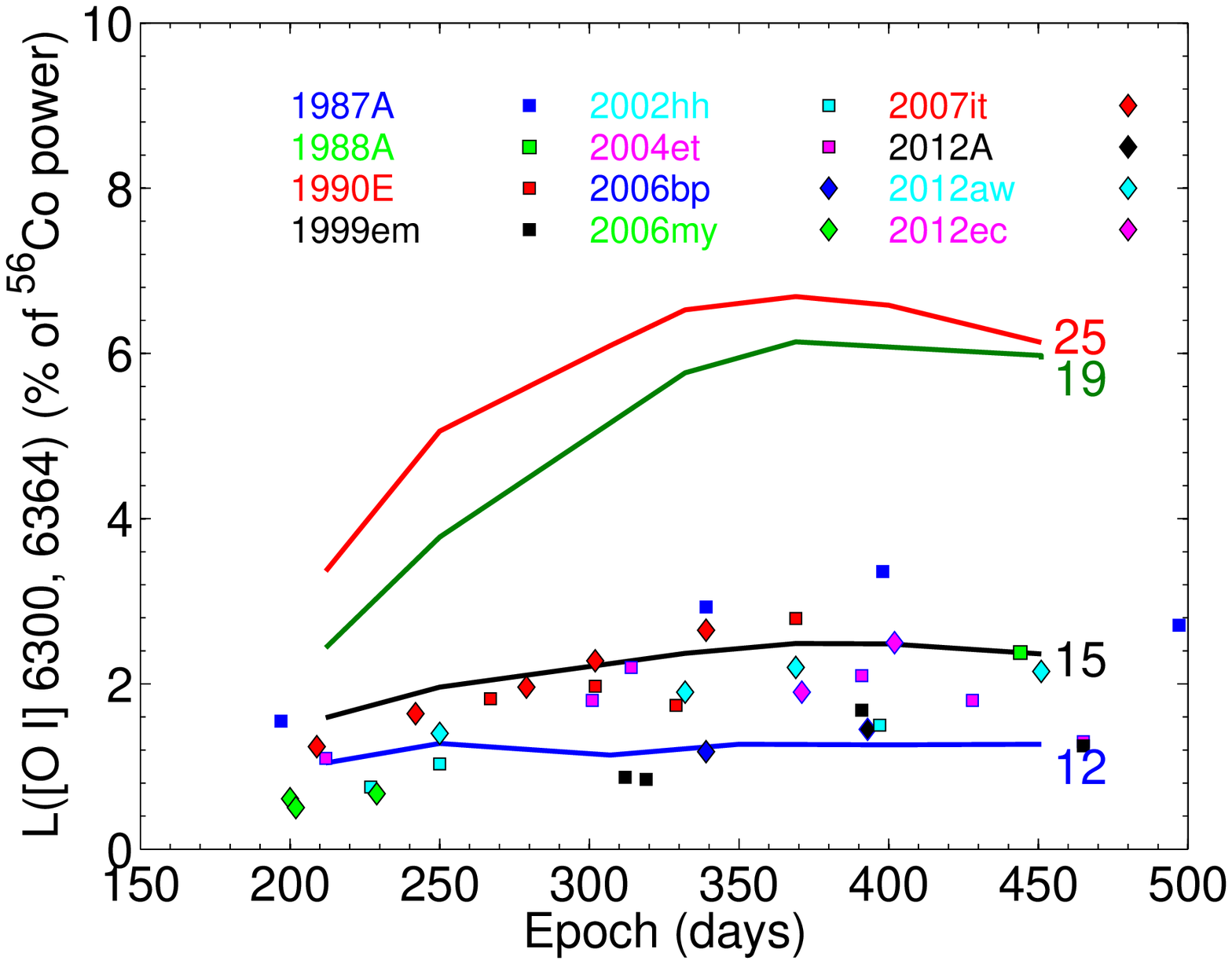}
\caption{Model tracks for luminosities in [O I] 6300, 6364 for 4 different progenitors ($M_{ZAMS}=12,15,19,25\ M_\odot$), compared to a sample of 12 Type II SNe. From \citet{J15b}.}
\label{fig:oItracks}
\end{figure}

The [O I] 6300, 6364 lines are efficient cooling lines and typically reemit a large fraction of the heating of the oxygen-zone layers. Their strengths are thus indicators of the oxygen mass, which from stellar evolution models is strongly dependent on the progenitor mass. Model luminosity tracks for different $M_{ZAMS}$ are presented in \citet{Jerkstrand2012,Jerkstrand2014} (see also Fig. \ref{fig:oItracks}). The method is particularly useful for observations around a year after explosion, when the [O I] lines become optically thin, temperatures and densities are in a regime favorable to [O I] 6300, 6364, and we are confident that $M_{O} \approx M_{O I}$. The current picture points to relatively limited amounts of oxygen produced in Type IIP SNe, and an origin in $M_{ZAMS} \lesssim 20$ \msun\ stars (Fig. \ref{fig:oItracks}). Another method to determine progenitor masses using the widths of the [O I] lines has been proposed by \citet{Dessart2010}. 

The luminosity of the O I doublet depends in general on $M_{\rm OI}, e^{-\Delta E/T(t)}$, and $n_e(t)$ in the optically thin NLTE phase (Sect. \ref{sec:NLTE}). LTE is valid until quite late times removing the $n_e$ dependency, but the exponential dependency on $T(t)$ makes it difficult to make any meaningful estimates of $M_{\rm OI}$ by inverse analytic modelling. Between $T=3000$ K and $T=6000$ K, the inferred O I mass changes by a factor of 45! Thus, one needs strong constraints on the temperature. One approach is to use the [O I] 5577 / [O I] 6300, 6364 ratio as a thermometer. The measurement of [O I] 5577 is intricate as it is weak in Type II SNe, but the method has been demonstrated to be feasible and in good agreement with forward modelling \citep{Jerkstrand2014}.

Few other oxygen lines are distinct in Type II SNe. One exception is O I 1.13 $\mu$m which is often observed to be strong, well above any plausible recombination luminosity. Its high strength likely arises as a fluorescence effect when Ly$\beta$ photons are absorbed in O I 1025, a process that gives indication of mixing of O and H clumps on small scales \citep{Oliva1993}.

\subsubsection{Calcium lines} 
\index{Calcium lines, nebular phase}
\citet{Kirshner1975}, and later \citet{Li1993}, demonstrated that the [Ca II] 7291, 7323 lines in H-rich SNe arise mainly from primordial (solar-abundance) Ca in the H-zone, and not from the synthesized calcium. The reason is the much larger mass of the H zone ($\sim$ 10 \msun) compared to the Si/S/Ca zone ($\sim 0.1$ \msun), which leads to more energy being reprocessed there. In addition, [Ca II] 7291, 7323 is a very efficient cooling channel that emits a large fraction of deposited energy even when the calcium abundance is low (solar). Thus, the [Ca II] emission lines
probe mainly the amount of energy processed by the hydrogen gas, and one expects them to be
stronger for higher H-zone masses and for closer mixing with \ni56.
\citet{Kozma1998b} could confirm the dominance of H-zone emission of [Ca II], but pointed out that
a contribution by the O-zone can occur in models using efficient convection and overshooting, where some Ca has been mixed out into the O layers.

Both [Ca II] 7291, 7323 and the Ca II NIR triplet are initially formed by thermal collisional excitation.
The Ca II NIR / [Ca II] 7291, 7323 ratio decreases with time as the temperature decreases. At later times
fluorescence following UV pumping in the Ca II HK lines becomes important. \citet{Li1993} estimate an epoch of $\sim$350d for this, while \citet{Kozma1998b} finds 500-800d in their models. Fluorescence takes over in the triplet first, roughly when $T \lesssim$ 5000 K. In the limit that both [Ca II] 7291, 7323 and Ca II NIR are driven by HK pumping, their intensity ratios will approach unity. 

The high optical depth of the triplet lines means that the Ca II 8498 line can scatter in the Ca II 8542 line (1500 \kms\ separation), and the Ca II 8542 line can scatter in the Ca II 8662 line (4200 \kms\ separation). To what extent this happens depends on the distribution of the H-zone gas and its temperature. In Figure \ref{fig:IIPmodel} (top) it can be seen that full scattering to the 8662 line has occurred in the model,
whereas the observed spectrum shows a distinct 8542 line that has not scattered. One should also note that O I 8446 will scatter in the Ca II NIR triplet, and [C I] 8727 can blend with its red wing. 

\subsection{Stripped-envelope SNe}
\index{Type Ib/c SNe, nebular phase}
\index{Type IIb SNe, nebular phase}
\index{Stripped-envelope SNe, nebular phase}

The first stripped-envelope nebular spectral models were calculated by \citet{Fransson1989}, of 4 and 8 \msun~He core explosions (Type Ib SN).
The models produced emission lines of Mg I] 4571, [O I] 5577, Na I D, [O I] 6300, 6364, O I 7774, [Ca II] 7300, Ca II NIR + [C I] 8727, as typically observed. 
The models were important in self-consistently predicting the range of temperatures and ionization expected, $T = 2000-8000$ K and $x_e \sim 0.1$. It was shown that an assumption of strong microscopic mixing of the SN ejecta gives model spectra discrepant with observations, consistent with the expected inefficiency of atomic diffusive mixing (Section \ref{sec:mixing}). The implication is that some caution is needed in interpreting single-zone models for core-collapse SNe,
and for highest accuracy multi-zone explosion modelling is needed.


Multi-zone spectral models of Type IIb SNe (only small amounts of hydrogen left) have been computed by \citet{Houck1996}, \citet{Maurer2010}, and \citet{J15a}. These models were used to study the behavour of H$\alpha$ from the hydrogen shell, the emission from the helium envelope, and the oxygen line brightness. The SNe modelled so far show nucleosynthesis pointing to an origin in low or intermediate mass stars, where the hydrogen envelope stripping must have occurred by Roche lobe overflow to a companion. Problems to explain apparent late-time H-alpha emission appear to have been resolved by including [N II] 6548, 6583 emission from the helium envelope \citep{J15a}, which mimics broad H$\alpha$ emission.

The first nebular Type Ic models were calculated by C. Kozma and presented in \citet{Sollerman2000}. These were 1D models of energetic explosions of 6 and 14 \msun~CO cores, with the aim to model SN 1998bw - the first SN associated with a gamma-ray burst. As for other SN classes, 1D explosion models gave poor reproduction of observed line profiles, demonstrating that some kind of mixing or asymmetric
explosion occurs also in Type Ic SNe. The model set was important in showing that models that give
good fits to early-time light curves and spectra can still be rejected from late-time comparisons. 

\begin{figure}
\centering
\includegraphics[width=0.48\linewidth]{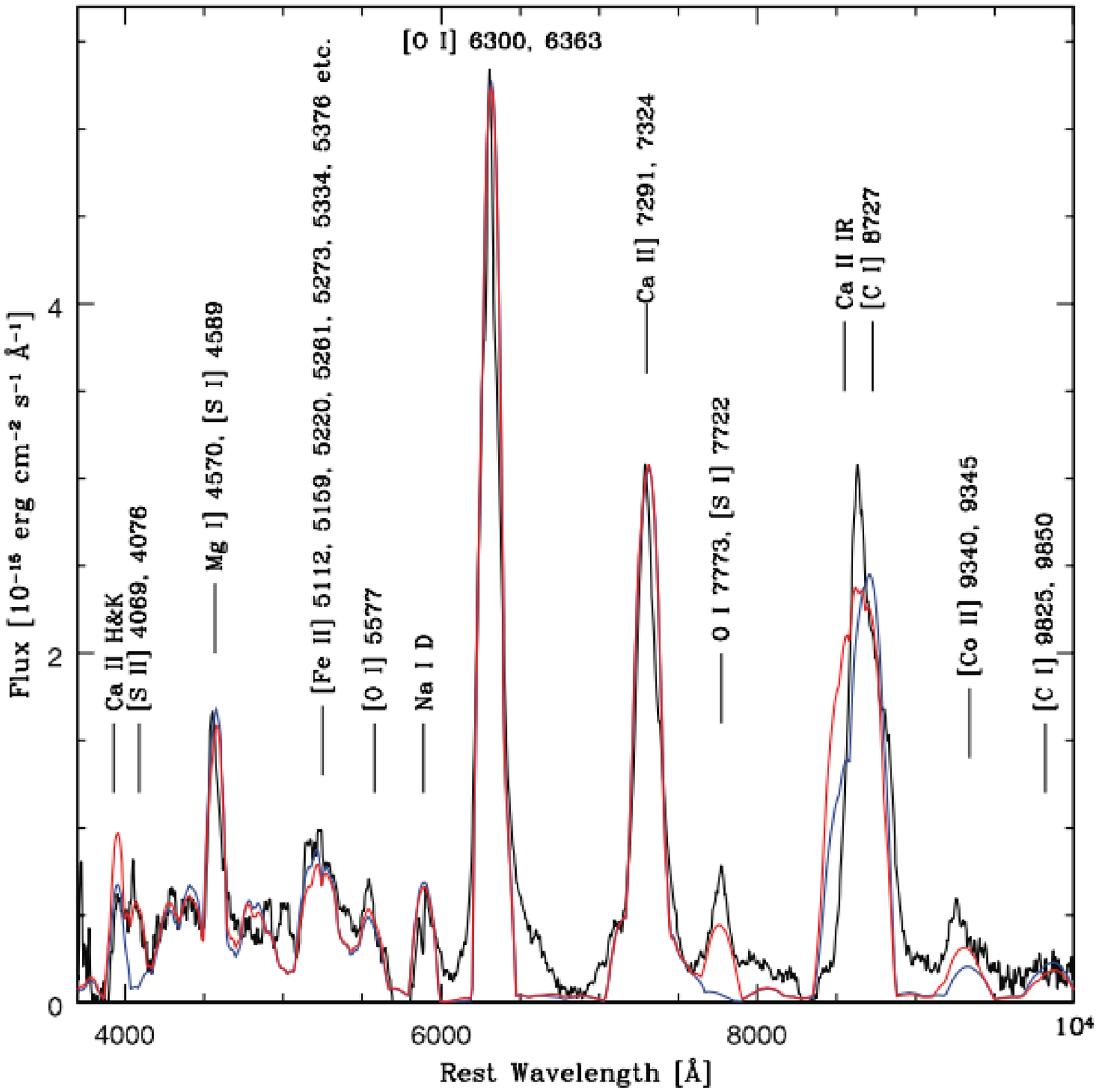}
\includegraphics[width=0.48\linewidth]{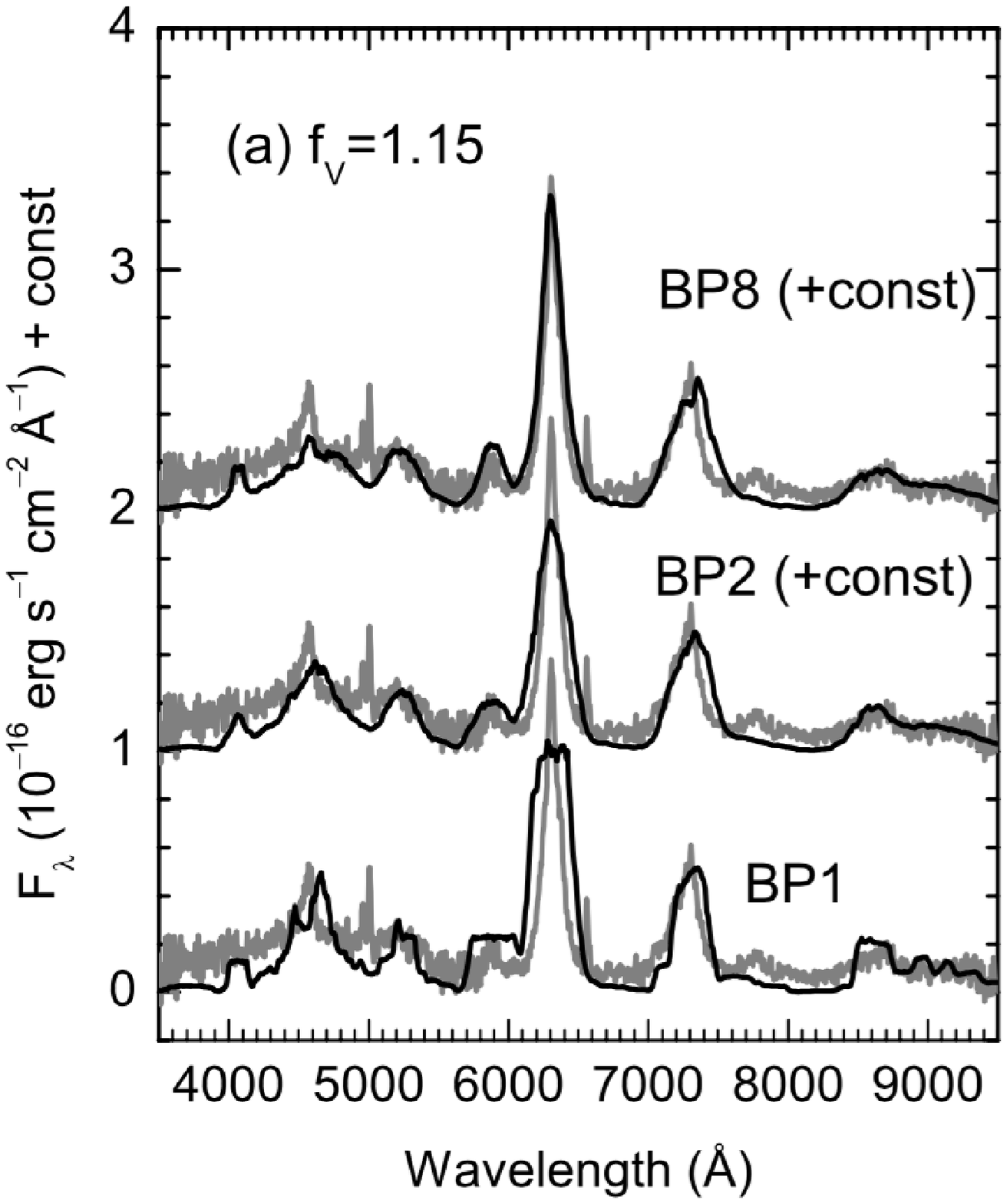}
\caption{Examples of stripped-envelope models. \emph{Left:} A single-zone Type Ic model compared to SN 2007gr \citep{Mazzali2010}. \emph{Right:} A 2D Type Ic model compared to SN 1998bw \citep{Maeda2006a}. The effects of increasing the explosion asymmetry (BP value) are illustrated.}
\label{fig:Ic}
\end{figure}

A single-zone approach has been taken in a series of papers by P. Mazzali \citep[e.g.][]{Mazzali2001, Mazzali2004, Mazzali2010} (see Figure \ref{fig:Ic} for an example). The majority of SNe analysed with single-zone models have oxygen mass estimates of 0.5-1.5 \msun, which suggests a low/intermediate progenitor mass range.
Some show only a few tenths of solar mass of oxygen \citep[e.g.][]{Sauer2006}, indicating that Nature can produce bare CO cores of as low mass as 2 \msun.
The Axelrod method has also been implemented in 2D \citep{Maeda2006a}.  2D explosion models offer a way to reproduce narrow lines from intermediate-mass element and broad lines from iron-group elements, as observed in SN 1998bw, if the viewing angle is close to pole-on (Figure \ref{fig:Ic}, right). 


Below some further comments are given on the formation of lines of magnesium, oxygen, and calcium, which are typically strong in stripped-envelope SN spectra.

\subsubsection{Magnesium lines}
\index{Magnesium lines, nebular phase}
The most prominent Mg line is Mg I] 4571, which is the first transition in Mg I.
Models show that the neutral fraction is typically small ($x_{\rm MgI} \sim 10^{-3}$) due to efficient photoionization both from ground state and excited states \citep{J15a}. This puts the line formation in a regime where both cooling and recombination can be important. The luminosity can show a dramatic increase at density thresholds above which cooling takes over. The line may be affected by line blocking to a significant extent, the efficiency of which increases rapidly blueward of 5000 \AA. This produces an asymmetric line profile with a blue-shifted peak.

Another distinct line produced that has been observed in many SNe is Mg I 1.504 $\mu$m. This is a pure recombination line,
free of line blocking. As models often give $M_{Mg} \approx M_{Mg II}$, this line becomes proportional to the magnesium mass, and can be used as a diagnostic of this \citep{J15a,J17}.



\subsubsection{Oxygen lines}

The [O I] 6300, 6364 doublet is, as in Type II SNe, an important coolant of the oxygen layers
and therefore a good diagnostic of their mass. Due to the higher expansion velocities in stripped-envelope SNe compared to H-rich SNe, the lines enter the optically thin regime at an earlier epoch, with $t_{\tau 6300}=1$ at 360d for $M=1$, $f=0.1$ and $V=3500$ \kms (using Eq. \ref{eq:thin}). Because the core expansion is faster than the line separation between the 6300 and 6364 lines (3047 km s$^{-1}$) the lines are blended and the line ratio has to be estimated by fitting the single blended feature.

[O I] 5577 is often distinct early on, and the [O I] 5577 / [O I] 6300, 6364 ratio may be used as a thermometer. This method breaks down quite early as [O I] 5577 falls out of LTE. This occurred after $\sim$150d in a model grid of Type IIb SNe \citep{J15a}. The [O I] 5577/[O I] 6300, 6364 ratio has been shown to depend on clumping \citep[e.g.][]{Maurer2010}, and holds some promise to be used as diagnotic for this.

When O I 7774, O I 9264, O I 1.129+1.130 $\mu$m and O I 1.316 $\mu$m are formed in the recombination regime,
their strengths 
may be used to estimate the quantity $n_e f^{1/2}$, under the assumption that $n_{\rm OII} \sim n_{\rm e}$. Effective recombination rates to be used for this have been calculated by \citet{Maurer2010} and \citet{J15a}. However, models also demonstrate complications and deviation from this regime. The meta-stable behaviour of many excited states in O I leads to significant optical depths for hundreds of days, and this produces a scattering contribution to the lines. At high densities cooling can occur in the O I 7774 transition \citep{Maurer2010}, significantly boosting it over its recombination luminosity.
While O I 7774 is relatively free from contaminations, O I 9263, O I 1.13 $\mu$m and O I 1.31 $\mu$m can be significantly blended with other lines \citep{J15a}.


\subsubsection{Calcium lines}
[Ca II] 7291, 7323 is typically the main coolant of the explosive oxygen burning ashes, and the calcium line
strengths therefore relate to how much energy is reprocessed by these layers. Standard explosion models ejecting $\sim$0.1 \msun\ of Si/S/Ca material have been shown to give good agreement with Type IIb SNe \citep{J15a}.

The Ca II NIR lines, other the other hand,
are after $\sim$200d formed mainly by fluorescence following HK absorption, mainly in the $^{56}$Ni ashes.

The [O I] 6300, 6364 / [Ca II] 7291, 7323 ratio is often used in the literature as a diagnostic of core mass, but there are
many issues with this method. The calcium emission in stripped-envelope SNe comes from an explosively made region, whose size and
distribution depend on the explosion energy. As long as the link between progenitor mass and explosion energy is unknown,
a diagnotic involving calcium is not robustly linked to core mass. 

\begin{figure}
\centering
\includegraphics[width=0.7\linewidth]{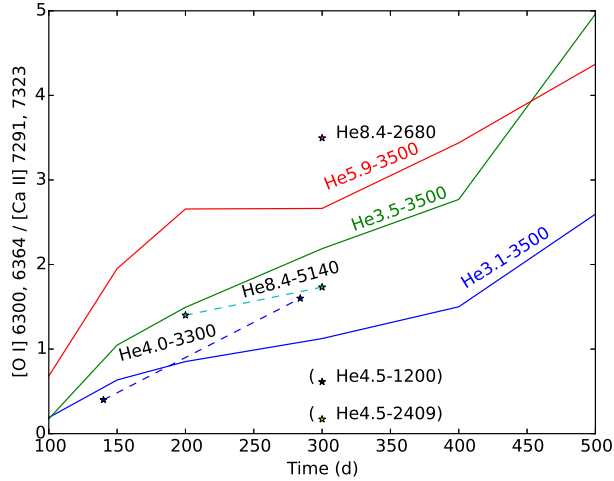}
\caption{Model predictions of the [O I] 6300, 6364 / [Ca II] 7291, 7323 ratio in stripped-envelope SNe. The He8.4 and He4.5 models are from \citet{Fransson1989}. The He4.0 model is from \citet{Houck1996}. The He3.1, He3.5 and He5.9 \msun\ models are from J15a. The grid
demonstrates complex dependency on He core mass, velocity, time, as well as nucleosynthesis model.}
\label{fig:OICaII}
\end{figure}

Figure \ref{fig:OICaII} shows model tracks of the [O I] 6300, 6364 /[Ca II] 7392, 7323 ratio from the models of \citet{Fransson1989}, \citet{Houck1996} and J15a. The grids indicate increase of this ratio with helium core mass, for other parameters fixed (and keep the caveat mentioned above in mind). The \citet{Fransson1989} grid shows a decrease of the ratio with higher core velocity, and all grids shows strong time dependency. As discussed in \citet{Fransson1989}, the very low ratios for the 4.5 \msun\ He cores are due to a large fraction of calcium in the O/Mg zone, something seen more rarely in modern explosion models. They are therefore marked in parenthesis.

\subsection{Thermonuclear supernovae} 
\index{Type Ia SNe, nebular phase}
The first Type Ia nebular models were computed by \citet{Axelrod1980}. These models are single-zone, pure iron-group composition, with relatively simple physics. 
The model spectra nevertheless showed good resemblance with observed Type Ia spectra,
strengthening the exploding white dwarf idea to explain Type Ia SNe. In particular were the strongest observed lines reproduced by Fe II emission at 5200 and 7155 \AA, Fe III emission at 4700 \AA, and Co III emission at 5900 \AA. 
The cobalt line evolution showed consistency with a declining abundance of cobalt, providing strong
support for the $^{56}$Co decay model.
Axelrod could also show that the ``quasi-continuum'' between the strong lines was formed by the
overlap by a large number of weaker spectral lines, not needing any true continuum emission.
Axelrod's technique, albeit without the global radiative transport component, has been adapted and applied to analyze Type Ia spectra in many subsequent papers. Single-zone models likely work reasonably well due to the homogenous composition of Type Ia SNe. \citet{RuizLapuente1992} developed a method to use Type Ia nebular spectra to determine the extinction towards the SN.

Multi-zone nebular models of the fast deflagration simulation W7 of \citet{Nomoto1984} have been presented in several papers \citep{RuizLapuente1995,Liu1997a,Sollerman2004,Leloudas2009,Maeda2010,Maurer2011,Mazzali2011,Fransson2015} using a variety of codes. \citet{Liu1997a} showed that the very high $^{58}$Ni abundance in W7 (Ni/Fe = 4 times solar) gave a much too strong [Ni II] 7378 line,
and SNe with MIR data show the same discrepancy for [Ni II] 6 $\mu$m \citep{Leloudas2009}.
However, \citet{RuizLapuente1992} estimate a ratio of 3 in SN 1995G, suggesting some variety.
\citet{Liu1997a} show how the ionization state and temperature increases with velocity coordinate (to about 8000 \kms), varying from FeI+FeII-dominated mix and $T\sim 3000$ K at the centre to Fe III + Fe IV dominated mix and $T\sim 7000$K at 8000 \kms, at 300d. \citet{Liu1998} discusses the physical ionization mechanisms at late times, finding non-thermal ionization and charge transfer to be more important processes than photoionization. Models generally show continuously decreasing ionization with time, with Fe I becoming the dominant ion after about two years \citep{Sollerman2004,Fransson2015}. Even before this time, Fe I plays an important role in the optical radiative transfer \citep{Axelrod1980}.

\begin{figure}[htb]
\centering
\includegraphics[width=0.8\linewidth]{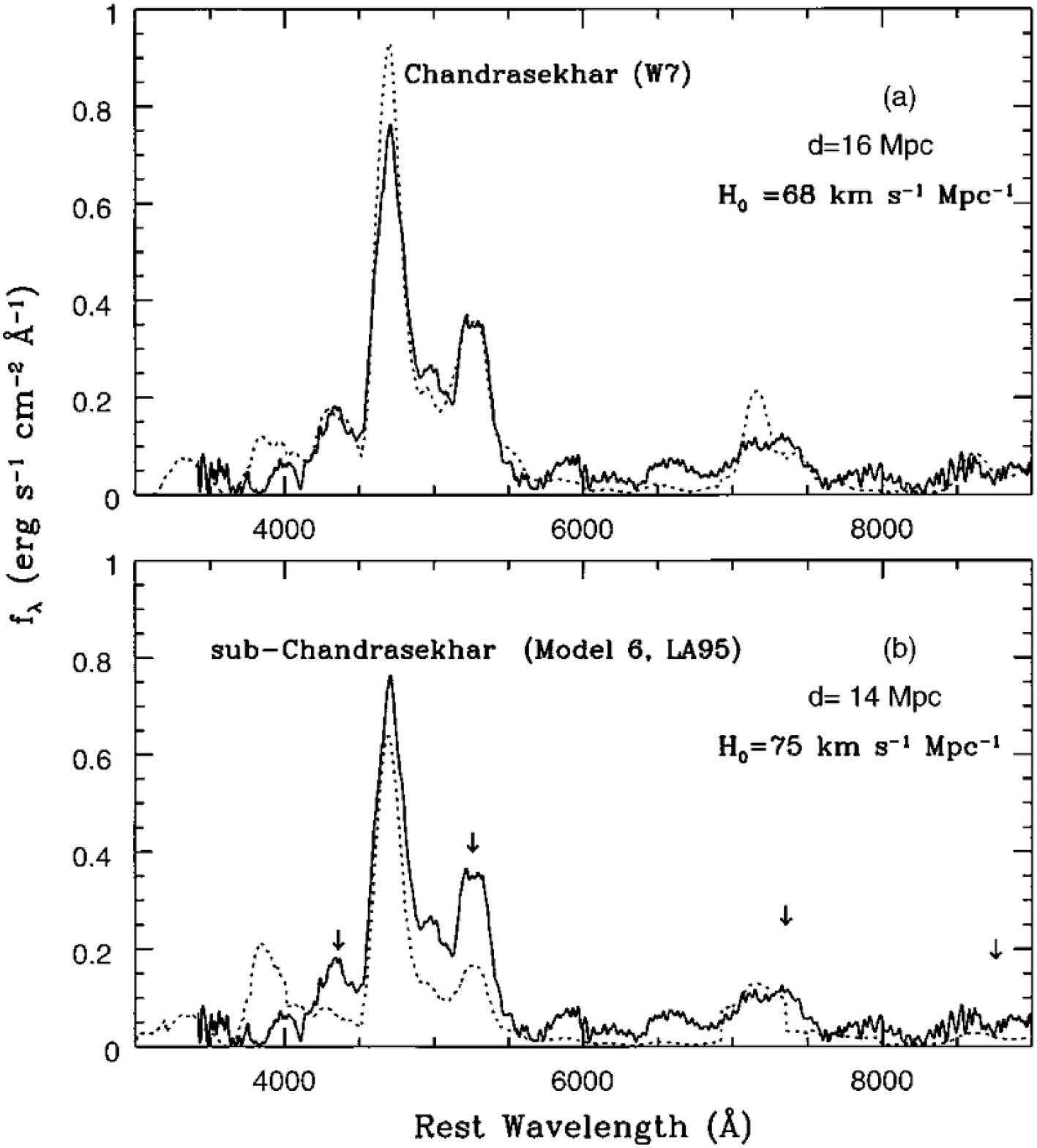}
\caption{Examples of Type Ia model spectra, from \citet{RuizLapuente1996}. Chandra (top) and sub-Chandra (bottom) models compared to SN 1994D. See also \citet{Mazzali2015} for a model with line IDs.}
\label{fig:Ia2}
\end{figure}

\citet{RuizLapuente1996}, \citet{Liu1997a}, and \citet{Mazzali2015} computed sub-Chandrasekhar models, pointing out that they are hotter and more ionized than W7 in the central regions, producing somewhat higher Fe III/Fe II line ratios. They also have weaker [Ni II] lines due to the smaller amount of $^{58}$Ni. \citet{Liu1997a} found a better fit for the sub-Chandra model to a series of observed SNe, whereas \citet{RuizLapuente1996}
favoured the W7 model for SN 1994D. \citet{Eastman1993} and \citet{Liu1997b} computed spectra of delayed detonation model DD4 of \citet{Woosley1994}. Pure deflagration models were studied by \citet{Kozma2005}, who found strong O I and C I lines in these to be inconsistent with observations. Two examples of spectral models are shown in Fig. \ref{fig:Ia2}.

For the first few hundred days, the ejecta are hot and ionized. Most radioactive energy is converted to heat, and is re-emitted in the optical and NIR. The spectrum is dominated by efficient cooling lines of iron, cobalt, and nickel. At later times, the ejecta pass through a rapid phase of cooling as the cooling switches from excited multiplets to ground multiplet transitions. The bulk of SN emission then moves to the mid-infrared, an effect dubbed the infrared catastrophy. The optical and near-infared regions then become dominated by fluorescence of UV emission \citep{Fransson2015}, and an accurate radiative transfer treatment is necessary. The fluorescence maintains optical output at about 20\% of the bolometric luminosity.

\section{Conclusions}
Supernovae in the nebular phase provide a wealth of information about the interiors of the exploded stars. 
From the first simple models devised some 35 years ago, we today have sophisticated codes capable of testing stellar evolution, explosion, and nucleosynthesis models in high level of detail. The nucleosynthesis of isotopes such as oxygen, magnesium, and nickel can be estimated, providing constraints on both progenitors and explosion physics. 
Some signatures need advanced models for interpretation, whereas some can be analyzed with simpler analytic approaches.
The vast majority of core-collapse SNe show nucleosynthesis consistent with an origin as $M_{ZAMS} \lesssim 20$ \msun\ stars,
and there may be a shortage of events from more massive progenitors compared to standard IMF expectations.
Line profiles may be used to probe the morphology of the ejecta for the various elements, which puts multi-D hydrodynamic models to the test. Line luminosities provide information about masses, emitting volumes, and physical conditions.
The evolution of light curves provides results on masses of radioactive isotopes such as $^{56}$Co, $^{57}$Co and $^{44}$Ti. By continuing 
efforts at determining the nucleosynthesis and inner ejecta morphology of SNe of different types much remains still to be learned about supernovae and the origin of the elements.

\begin{acknowledgement}
I would like to thank J. Spyromilio, C. Fransson, K. Maeda, R. McCray, S. Taubenberger, J. Sollerman, P. Mazzali, S. J. Smartt, M. Ergon, and P. Ruiz-Lapuente for useful comments on the manuscript.
\end{acknowledgement}
%
%
%

\section{Cross-references}
\begin{itemize}
\item Dust and molecular formation in supernovae
\item Spectra of supernovae during the photospheric phase
\item Nucleosynthesis in spherical explosion models of core collapse supernovae
\item The Multi-Dimensional Character of Nucleosynthesis in Core Collapse  Supernovae
\item Nucleosynthesis in Hypernovae: Gamma Ray Bursts 
\item Nucleosynthesis in Thermonuclear Supernovae
\end{itemize}

\bibliographystyle{plainnat}
\bibliography{ref_anders}
\end{document}